\documentclass{iopart}
\usepackage{graphicx}  
\usepackage{latexsym}
\usepackage{amssymb}
\usepackage{cite}

\jl{6}        
\eqnobysec    

\def\subsec#1{\vskip12pt\noindent{\textit{#1}}}

\def\Strut{\rule[-5pt]{0pt}{15pt}}
\def\SStrut{\rule[-7pt]{0pt}{19pt}}

\def\beq{\begin{equation}}
\def\eeq{\end{equation}}

\def\rmd{{\rm d}}

\DeclareSymbolFont{AMSb}{U}{msb}{m}{n}
\DeclareMathSymbol{\R}{\mathbin}{AMSb}{"52}
\DeclareMathSymbol{\I}{\mathbin}{AMSb}{"49}
\DeclareMathSymbol{\C}{\mathbin}{AMSb}{"43}

\def\selfd{\,\mathop{\tilde{\,}}\nolimits\,}
\def\leftselfd{\selfd\kern-1.7pt}

\begin{document}

\title
[General relativistic Poynting-Robertson effect II]
{The general relativistic Poynting-Robertson effect
II: A photon flux with nonzero angular momentum}

\author{
Donato Bini${}^*{}^\S{}^\ddag$, Andrea Geralico${}^\S{}^\sharp$, 
Robert T. Jantzen${}^\P{}^\S$ 
Old\v{r}ich Semer\'ak${}^\mathdollar$
and 
Luigi Stella$^\dag$
}
\address{
  ${}^*$\
Istituto per le Applicazioni del Calcolo ``M. Picone'', CNR I-00161 Rome, Italy
}
\address{
  ${}^\S$\
  International Center for Relativistic Astrophysics - I.C.R.A.,
  University of Rome ``La Sapienza'', I-00185 Rome, Italy
}
\address{
  ${}^\ddag$\
  INFN sezione di Firenze, I--00185
  Sesto Fiorentino (FI), Italy
}
\address{
  $^\sharp$
  Physics Department,
  University of Rome ``La Sapienza,'' I--00185 Rome, Italy
}
\address{
  ${}^\P$\
Department of Mathematical Sciences, Villanova University, Villanova, PA 19085,  USA
}
\address{
  ${}^\mathdollar$\
Institute of Theoretical Physics, Faculty of Mathematics and Physics, 
Charles University in Prague, Czech Republic
}
\address{
  ${}^\dag$\
Osservatorio Astronomico di Roma, via Frascati 33, I-00040 Monteporzio Catone (Roma), Italy
}

\begin{abstract}
We study the motion of a test particle in a stationary, axially and 
reflection symmetric spacetime of a central compact object, as affected by 
interaction with a test radiation field of the same symmetries. 
Considering the radiation flux with fixed but arbitrary (non-zero) angular 
momentum, we extend previous results limited to an equatorial motion 
within a zero-angular-momentum photon flux in the Kerr and Schwarzschild 
backgrounds. While a unique equilibrium circular orbit exists if the photon flux 
has zero angular momentum, multiple such orbits appear if the photon 
angular momentum is sufficiently high.
\end{abstract}

\pacno{04.20.Cv}

\section{Introduction}

Calculating the motion of test particles in black hole and other compact object fields is a standard way to reveal the properties of strong field spacetimes, and possibly also to estimate the behavior of matter in certain actual astrophysical systems. While various simple limiting cases which cover just some aspects of the problem --- the motion of free particles in particular --- now belong to textbook exercises, it is clear that the real picture is much more complicated due to various ``perturbations'' both  of the ``background'' as well as of the ``particle.'' Even without taking into account any quantum effects, extra dimensions, non-standard topology, additional hypothetical fields or a more sophisticated cosmological setting, there still remain several obvious effects whose relative importance and interconnections are not clear. First the compact objects discussed and modeled in astrophysics are {\em observable} and hence {\em interacting}, so some matter must be present around them. 
When studying the behavior of a certain ``particle'' of that matter, the overall gravitational and also direct physical influence of the ``bulk'' material should be taken into account. Due to the very energetic processes occurring in strong and non-homogeneous fields close to compact objects, the matter is highly ionized and prone to electromagnetic interaction, either with the field generated by its own currents or (in the case of neutron stars) with that maintained by the central body. High energies and electromagnetic fields in turn produce radiation --- in fact {\em very intense} (and hard) radiation, namely that which enables us to detect the source. Also there are more ``hairs'' (besides just mass) on the side of the test body whose motion is in question. The latter is often endowed with proper angular momentum at least (``pole-dipole approximation''), and probably even higher multipoles may be relevant in some cases. To appreciate this, one should note that the ``particle'' may sometimes represent such a large and elastic body as a whole star (as in the discussion of motion around supermassive black holes in galactic nuclei).

When asking about the short term evolution of a body located deep in a well of the exterior field of a black hole, the above influences may generically be neglected, because the force field is typically dominated by the center's gravitational pull there. However, when studying the longer term evolution of the system, its stability comes into consideration and this may well be affected by various ``external'' ingredients, mainly at moderate distances from the center. It has in fact become clear that the Kerr(-Newman) family of black holes --- though being the most general among {\em isolated} and stationary cases in asymptotically flat universes --- is very special in its multipole structure, in particular, it is just the one required for uniqueness theorems that permits the solution of geodesic equations of motion in terms of separated first integrals (see e.g. \cite{Will-09}). This full integrability is lost if any of the assumptions (isolation, stationarity, asymptotic flatness) is removed --- and they are all violated in astrophysical circumstances. Then, however, even a tiny ``perturbation'' generically makes the dynamics of test motion chaotic, which means that the long term evolution of the system may depart considerably from that obtained in the Kerr(-Newman) background.

In the present series of articles, we also focus on just one of the above ``perturbations,'' namely on the effect of a test radiation field. There probably exists a complicated radiation situation around a real accreting compact object, but we will limit ourselves to the case of a coherent flux, composed of  photons traveling along geodesics in some preferred direction. Possible scenarios include a hot (young or accreting) neutron star, radiating roughly radially, a black hole {\em accreting} radiation roughly radially,\footnote
{In realistic accretion systems the radiation would hardly come to the hole in a spherically symmetric manner, but one may also refer here to a rather solitary hole in a ``bath" of relic radiation.}
or a system with an accretion disc which radiates preferentially in the perpendicular (``vertical") direction. It has been mainly this last case that attracted astrophysical attention, in connection with whether such a disc radiation could accelerate particles along the rotation axis (and thus help to form jets observed in many accretion systems) --- see e.g. \cite{VokrouhlickyK-91}. With similar motivation, the authors of \cite{KeaneBS-01} examined radial motion of a material particle in a Schwarzschild background, when subjected to a radial radiation field.

However, we will not limit ourselves necessarily to flux geometries with direct astrophysical justification here. Instead we will focus on a stationary axisymmetric beam of photons emitted from a ring in the equatorial plane where it impacts a test particle in motion, investigating the qualitative features of the resulting orbits as a function of the radiation strength and angular momentum. In particular we
 check whether this system allows for some particular, remarkable results of interplay between the radiative flux pressure and friction-like drag, the test particle inertia and the gravitational attraction (and dragging) of the central object.

In a previous article \cite{bijanste}, steps were already taken in that direction. Test particle motion in a Kerr spacetime was studied in its equatorial plane with a stationary axisymmetric test electromagnetic flux consisting of a beam of zero angular momentum photons moving in the purely radial outward direction in that plane with respect to the locally nonrotating frames naturally associated with the family of zero angular momentum observers (ZAMOs). 
Unless the radiation force is so strong 
or the test particle's initial energy is too high 
so that it escapes to infinity, 
there is always a critical radius at which test particles come to rest with respect to the ZAMOs due to the drag forces exerted by the radiation. This equilibrium situation represents a balance of the outward radiation force with the inward gravitational force.
Of course the problem of radiation drag exerted on a material particle is by no means new, an effect bearing the names of Poynting \cite{poynting} and Robertson \cite{robertson} who tackled it within Newtonian theory and within linearized general relativity, respectively. See \cite{bijanste} for a brief history of this effect with further references. All of these models involve a single coherent stream of photons emanating from a surface surrounding a central source. If the critical radius lies inside the cutoff radius of the emitting surface beyond which the model is no longer relevant, of course, test particles fall into this surface.

More realistic general relativistic models allow for photons to be emitted in any direction from the emitting region surface. Abramowicz, Ellis and Lanza \cite{abr-ell-lan} considered the special case of only radially moving test bodies in the Schwarzschild spacetime, and they classified quantitatively the types of motion which result, including the radial equilibrium solutions in which the test body remains at rest under the combined inward gravitational 3-force and the outward photon pressure. 
Miller and Lamb  \cite{lam-mil,mil-lam3,mil-lam1} generalized this to arbitrary test equatorial plane particle motion in the Schwarzschild spacetime and then considered the effects of a small rotation of the gravitational source, but did not study the equilibrium solutions in detail. 
This was instead done recently by Oh, Kim and Lee \cite{oh}  using the Miller-Lamb slow rotation model for the source of the gravitational field, finding similar behavior to Bini et al \cite{bijanste} for the corotating equilibrium orbits.
The difficulty with these models is that they are much more complicated to work with compared to the simpler model with a unidirected radiation stream impacting the test particle, and are limited to slow rotation effects.

In the present article, to get some rough idea of the effects of strong rotation, we generalize our previous discussion \cite{bijanste} of test particle motion in the equatorial plane of a Kerr spacetime under the influence of a radial photon flux  with zero angular momentum to the case of nonzero angular momentum and hence a nonzero azimuthal component of the photon 4-momentum.
This is motivated by the observation that radiation emitted at the surface of a fast rotating source would be expected to have angular momentum correlated with that rotation.
If a test particle impacted by this radiation does not escape to infinity because the outward radiation force is too strong or its initial energy is too high, instead of coming to rest with respect to the ZAMOs at a single critical radius which is a stable equilibrium for the radial motion as in the previous case, the critical circular orbits have a constant nonzero azimuthal velocity equal to the azimuthal velocity of the photons at the critical radius. As the photon angular momentum impact parameter increases, multiple such critical radii appear, leading to a much more complicated dynamical scenario.
As in the previous article, we first describe the dynamical equations and conditions for the critical orbits in the reflection symmetric plane in a general orthogonally transitive, stationary, axially and reflection symmetric spacetime, and then specialize them to the equatorial plane of the Kerr and Schwarzschild spacetimes, outside of some cutoff radius of the ring surrounding the central object at which the radiation is emitted.

\section{Stationary, axisymmetric and reflection-symmetric spacetimes} 

{\it Metric and fiducial observers}

The metric of an orthogonally transitive  stationary axisymmetric spacetime using  coordinates $\{t,r,\theta,\phi \}$ adapted to the spacetime symmetries has a line element of the form \cite{MTW}
\beq
\label{metr_gen}
\rmd s^2= 
  g_{tt} \rmd t^2 + 2 g_{t\phi} \rmd t \rmd \phi + g_{\phi\phi}\rmd \phi^2 
  + g_{rr}\rmd r^2 + g_{\theta\theta} \rmd \theta^2\ \,,
\eeq
where all the metric coefficients depend only on $r$ and $\theta$
and $\partial_t$ (timelike) and $\partial_\phi$ (spacelike, with closed orbits) are  (commuting)  Killing vectors. 
As is the case for 
the black hole spacetimes, we further require the metric to be reflection-symmetric with respect to the equatorial plane $\theta=\pi/2$.

The time coordinate lines, when timelike, are the world lines of the static observers. 
The zero angular momentum observer (ZAMO) family of fiducial observers has instead a 4-velocity field $n$ equal to the future-pointing unit normal to the time coordinate hypersurfaces $t=$constant, namely
\beq
\label{n}
n=N^{-1}(\partial_t-N^{\phi}\partial_\phi)\,,
\eeq
where $N=(-g^{tt})^{-1/2}$ and $N^{\phi}=g_{t\phi}/g_{\phi\phi}$ are the lapse function and only nonvanishing component of the shift vector field respectively. 
Discussion here is limited to those regions of spacetime where the time coordinate hypersurfaces are spacelike: $g^{tt}<0$, i.e, outside the outer horizon of black hole spacetimes.
The ZAMO relative velocity of the static observers following the time coordinate lines at constant azimuthal angle $\phi$ is $\nu_{\rm(s)} =g_{\phi\phi}^{1/2} N^{-1}N^{\phi}$, whose absolute value goes to 1 at the outer boundary of the ergosphere in black hole spacetimes. Inside this surface its reciprocal $\bar\nu_{\rm(s)} =1/\nu_{\rm(s)}$ is the relative azimuthal velocity of the orthogonal timelike direction within the Killing cylinder of the $t$-$\phi$ coordinate surface, which goes to zero at the horizon. For the nonnegative Kerr rotation parameter range $0\le a/M\le 1$ that we assume here,  $\nu_{\rm(s)}$ and $\bar\nu_{\rm(s)}$ are negative.

An orthonormal frame adapted to the ZAMOs  is given by
\beq\fl\quad
\label{zamoframe}
e_{\hat t}=n\,,\quad
e_{\hat r}=\frac1{\sqrt{g_{rr}}}\partial_r \equiv \partial_{\hat r}\,,\quad
e_{\hat \theta}=\frac1{\sqrt{g_{\theta \theta }}}\partial_\theta \equiv \partial_{\hat \theta}\,,\quad
e_{\hat \phi}=\frac1{\sqrt{g_{\phi \phi }}}\partial_\phi \,,
\eeq
with dual frame
\beq
\fl\quad
\omega^{{\hat t}}= N\rmd t\,,\quad \omega^{{\hat r}} =\sqrt{g_{rr}} \,\rmd r\,,\quad 
\omega^{{\hat \theta}}= \sqrt{g_{\theta \theta }} \,\rmd \theta\,,\quad
\omega^{{\hat \phi}}=\sqrt{g_{\phi \phi }}(\rmd \phi+N^{\phi}\rmd t)\,,
\eeq
corresponding to the re-expressed form of the line element
\beq
\rmd s^2 = -N^2\rmd t^2 +g_{\phi \phi }(\rmd \phi+N^{\phi}\rmd t)^2 + g_{rr}\rmd r^2 +g_{\theta \theta}\rmd \theta^2\,. 
\eeq

The accelerated ZAMOs are locally nonrotating in the sense that their vorticity vector $\omega(n)$ vanishes, but they have a nonzero expansion tensor $\theta(n)$ whose nonzero
components can be completely described by a shear vector
\beq
\label{exp_zamo}
\theta(n) = e_{\hat\phi}\otimes\theta_{\hat\phi}(n)
           +\theta_{\hat\phi}(n)\otimes e_{\hat\phi}
\,, \quad \theta_{\hat \phi}(n)^\alpha = \theta(n)^\alpha{}_\beta\,{e_{\hat\phi}}^\beta
\,.
\eeq 
Since the expansion scalar $\theta(n)^\alpha{}_\alpha$ is zero, the expansion and shear tensors coincide.

The nonzero ZAMO kinematical quantities (acceleration $a(n)=\nabla_n n$ and shear tensor) as well as the associated  Lie relative curvature vector \cite{mfg,idcf1,idcf2,bjdf} only have nonzero components in the $r$-$\theta$ 2-plane of the tangent space,
\begin{eqnarray}
\label{accexp}
\fl\quad
a(n) & = & a(n)^{\hat r} e_{\hat r} + a(n)^{\hat\theta} e_{\hat\theta}
 =\partial_{\hat r}(\ln N) e_{\hat r} + \partial_{\hat\theta}(\ln N)  e_{\hat\theta}
\,,
\nonumber\\
\fl\quad
\theta_{\hat\phi}(n) 
& = & \theta(n)^{\hat r}{}_{\hat\phi} e_{\hat r} + \theta(n)^{\hat\theta}{} _{\hat\phi}e_{\hat \theta} 
  = -\frac{\sqrt{g_{\phi\phi}}}{2N}\,(\partial_{\hat r} N^\phi e_{\hat r} + \partial_{\hat\theta} N^\phi e_{\hat \theta})
\,, 
\nonumber\\
\fl\quad
k_{(\rm lie)}(n)
& = & k_{(\rm lie)}(n)^{\hat r} e_{\hat r} + k_{(\rm lie)}(n)^{\hat\theta} e_{\hat\theta}
 = -[\partial_{\hat r}(\ln \sqrt{g_{\phi\phi}}) e_{\hat r} + \partial_{\hat\theta}(\ln \sqrt{g_{\phi\phi}})e_{\hat\theta}]
\,.
\end{eqnarray}
In the static limit $N^\phi\to0$, the shear vector $\theta_{\hat\phi}(n)$ vanishes. 
In the equatorial plane the above quantities have only a radial component, as summarized in  Table 1 below for the Kerr spacetime.

\subsec{Photons}

Let a pure electromagnetic radiation field with the same symmetry properties as the spacetime be superimposed as a test field on this gravitational background, with the energy-momentum tensor
\beq
\label{ten_imp}
T^{\alpha\beta}=\Phi^2 k^\alpha k^\beta\,, \qquad 
k^\alpha k_\alpha=0\,, \qquad
k^\alpha \nabla_\alpha k^\beta=0 
\,,
\eeq
where the 4-momentum field $k$ is assumed to be tangent to a family of affinely parametrized null geodesics which in the equatorial plane has $k^\theta=0$ because of the reflection symmetry, so that the photon orbits remain in that plane. 
The 4-momentum and unit vector direction of the relative velocity with respect to the ZAMOs are then respectively
\beq
\label{eq:phot}
k= E(n)[n+\hat \nu(k,n)], \qquad  
\hat \nu(k,n)=\sin \beta\, e_{\hat r}+\cos \beta\, e_{\hat \phi}\,,
\eeq
where 
\beq
E(n)= - k \cdot n 
=\frac{E+LN^\phi}{N}
    =\frac{E}{N}(1+bN^\phi)\,,\qquad 
b\equiv\frac{L}{E}
\eeq
is the relative energy of the photons, $E=-k_t>0$ is the constant conserved energy associated with the timelike Killing vector field, $L=k_\phi$ is the  constant conserved angular momentum associated with the rotational Killing vector field, and
\beq
\label{cosbeta}
\fl\qquad
\cos \beta 
= \frac{L N}{\sqrt{g_{\phi\phi}} \,(E+LN^\phi) }
=\frac{bN}{\sqrt{g_{\phi\phi}} \,(1+b N^\phi)}
=\frac{b E}{\sqrt{g_{\phi\phi}} \,E(n) }\,,
\eeq
which implies
\beq\label{tanbeta}
 N|b\tan\beta| =[g_{\phi\phi}(1+bN^\phi)^2-b^2N^2]^{1/2} \,.
\eeq
The constant $b=L/E$ is the photon impact parameter \cite{MTW,chandra,GRintro}, which can be re-expressed in terms of the photon angle by inverting Eq.~(\ref{cosbeta}),
\beq\label{bp}
 b = \frac{ \sqrt{g_{\phi\phi}} \cos\beta}{N(1- \nu_{\rm(s)}\cos\beta)} 
 = \frac{ \sqrt{g_{\phi\phi}} \cos\beta}{N(1- \cos\beta/\bar\nu_{\rm(s)})}
\,.
\eeq
The limit  $\cos\beta\to \bar\nu_{\rm(s)}$ 
(which for a black hole spacetime can only occur in the ergosphere where $|\bar\nu_{\rm(s)}|<1$) leads to $\pm b\to\infty$. 
Restricting to the photons with $E>0$ (those with $E<0$ can only exist inside ergosphere), we will assume $1+bN^\phi>0$ so that $E(n)>0$ and $k$ is a future-directed vector. 
Note that the value of $1+bN^\phi$ is also constrained by the requirement that $|\cos\beta|\leq 1$. 

The case $\sin \beta >0$ corresponds to outgoing photons (increasing $r$) and $\sin \beta <0$ to incoming photons (decreasing $r$). Although we are primarily interested in the outgoing case, we include the incoming case for completeness.
The case $\sin\beta=0$ of purely azimuthal geodesic motion of the photons can only take place at the null circular geodesic radii and no other, so we exclude it.

Since $k$ is completely determined,  the coordinate dependence of the quantity $\Phi$  then follows 
from the conservation equations  $T^{\alpha\beta}{}_{;\beta}=0$, and will only depend on $r$ in the equatorial plane due to the axial symmetry.
From Eq.~(\ref{ten_imp}) using the geodesic condition for $k$, these can be written as
\beq
\label{flux_cons}
0=\nabla_\beta (\Phi^2 k^\beta)
=\frac{1}{\sqrt{-g}}\partial_\beta (\sqrt{-g}\,\Phi^2 k^\beta)
=\partial_r(\sqrt{-g}\,\Phi^2 k^r)
\,, 
\eeq
leading to 
$\sqrt{-g}\,\Phi^2 k^r
=N E(n)\sqrt{g_{\theta\theta}g_{\phi\phi}} \,\Phi^2\sin\beta
=\hbox{\rm const} = E \Phi_0^2 $ 
and therefore
\beq
\Phi^2=\frac{\Phi_0^2}{\sqrt{g_{\phi\phi}}\, N|b \tan \beta|} \,.
\eeq
Note that from Eq.~(\ref{tanbeta}) this expression has the $b\to0$ limit $\Phi_0^2/ \sqrt{g_{\theta\theta} g_{\phi\phi}} $ for photons in radial motion with respect to the ZAMOs \cite{bijanste}. 

The test radiation field is assumed to start at some axisymmetric emission surface which intersects the equatorial plane at the radius $r=R$, where $R$ is certainly greater than the horizon in a black hole spacetime. Abramowicz et al \cite{abr-ell-lan} choose $R/M =3$ to be the radius of the last circular null geodesic in the Schwarzschild spacetime, for example. This should be kept in mind when the region outside the outer horizon of black hole spacetimes is considered below.

\subsec{Test particle}

Consider now a test particle moving in the equatorial plane, i.e., with 4-velocity and 3-velocity with respect to the ZAMOs respectively
\beq\label{polarnu}
\fl\quad 
U=\gamma(U,n) [n+ \nu(U,n)]\,,\quad
\nu(U,n)\equiv \nu^{\hat r}e_{\hat r}+\nu^{\hat \phi}e_{\hat \phi} 
=  \nu \sin \alpha\, e_{\hat r}+ \nu \cos \alpha\, e_{\hat \phi}  \,,
\eeq
where $\gamma(U,n)=1/\sqrt{1-||\nu(U,n)||^2}$ is the Lorentz factor and the abbreviated notation $\nu^{\hat a}=\nu(U,n)^{\hat a}$ has been used. In a similarly abbreviated notation, $\nu = ||\nu(U,n)||\ge0$ and $\alpha$ are the magnitude of the spatial velocity $\nu(U,n)$ and its
polar angle  measured clockwise from the positive $\phi$ direction in the $r$-$\phi$ tangent plane.
Note that $\sin\alpha=0$ (i.e., $\alpha=0,\pi$) corresponds to purely azimuthal motion of the particle with respect to the ZAMOs, while $\cos\alpha=0$
(i.e., $\alpha=\pm \pi/2$) corresponds to (outward/inward) purely radial motion with respect to the ZAMOs.

Using the expression (\ref{n}) for $n$ leads to the coordinate components of $U$
\begin{eqnarray}
\label{Ucoord_comp}\fl\qquad
&& U^t\equiv \frac{\rmd t}{\rmd \tau}=\frac{\gamma}{N}, \qquad U^r\equiv \frac{\rmd r}{\rmd \tau}
=\frac{\gamma \nu^{\hat r}}{\sqrt{g_{rr}}}\,, 
\nonumber \\ \fl\qquad
&& U^\theta\equiv \frac{\rmd \theta}{\rmd \tau}=0, \qquad U^\phi\equiv \frac{\rmd \phi}{\rmd \tau}
=\frac{\gamma \nu^{\hat \phi}}{\sqrt{g_{\phi\phi}}}  -\frac{\gamma N^\phi}{N}
=\frac{\gamma }{\sqrt{g_{\phi\phi}}} \left( \nu^{\hat \phi}-\nu_{\rm(s)}\right)
\,,
\end{eqnarray}
where $\tau$ is a proper time parameter along $U$. 
Solving these for the polar angle and speed leads to
\begin{eqnarray}
\tan \alpha &=&\sqrt{ 
\frac{g_{rr}}{g_{\phi\phi}}
}\, 
\frac{\rmd r}{\rmd t}\left( \frac{\rmd \phi}{\rmd t}  +N^\phi\right)^{-1}\,, 
\nonumber \\
\phantom{xxx}\nu&=& 
\frac{1}{N}\sqrt{g_{rr} \left( \frac{\rmd r}{\rmd t} \right)^2
         + g_{\phi\phi} \left( \frac{\rmd \phi}{\rmd t}+N^\phi \right)^2} \,.
\end{eqnarray}

For geodesic motion the Killing energy $E_{\rm(p)} = -U_t$ and angular momentum $L_{\rm(p)} = U_\phi$ are conserved, and their ratio is the test particle impact parameter
\cite{MTW,chandra,GRintro}
\beq\label{bpp}
 b_{\rm(p)} = \frac{L_{\rm(p)}}{E_{\rm(p)}}
 = \frac{ \nu\sqrt{g_{\phi\phi}} \cos\alpha}{N(1- \nu_{\rm(s)}\nu\cos\alpha)} \,,
\eeq
reducing to the previous formula for photons when $\nu=1$ and $\alpha\to\beta$. 
This quantity is also constant for accelerated but circular orbits (where $\cos\alpha=\pm1$) at a constant radius and signed azimuthal velocity $\nu^{\hat\phi}=\nu\cos\alpha$, in which case this relation becomes
\beq\label{bppcirc}
 b_{\rm(p)} 
 = \frac{ \nu^{\hat\phi}\sqrt{g_{\phi\phi}} }{N(1 - \nu_{\rm(s)}\nu^{\hat\phi})} \,.
\eeq
If further one sets $\nu^{\hat\phi}=\cos\beta$, this reproduces the photon relation exactly, i.e., for circular motion at the same azimuthal velocity that the photons have, the photon and particle impact parameters coincide. This happens simply because the impact parameter is only a function of the azimuthal velocity.

\subsec{Radiation test particle interaction}

The scattering of the radiation from the test particle as well as the constant momentum-transfer cross section $\sigma$ are assumed to be independent of the direction and frequency of the radiation, characteristic of Thomson scattering. The associated force is then given by \cite{abr-ell-lan}
\beq
{\mathcal F}_{\rm (rad)}(U)^\alpha = -\sigma P(U)^\alpha{}_\beta \, T^{\beta}{}_\mu \, U^\mu \,,
\eeq
where $P(U)^\alpha{}_\beta=\delta^\alpha_\beta+U^\alpha U_\beta$ projects orthogonally to $U$.
The equation of motion of the particle then becomes
$ m a(U) = {\mathcal F}_{\rm (rad)}(U)$,
where $m$ is the mass of the particle and $a(U)=D U/\rmd\tau $ is its 4-acceleration.

To evaluate the radiation force which by definition lies in the local rest space of the test particle,
it is useful to introduce the relative decomposition of the photon 4-momentum $k$ with respect to the test particle 4-velocity in addition to the previous ZAMO decomposition,
\beq
\label{diff_obg}
k = E(n)[n+\hat \nu(k,n)]
= E(U)[U+\hat {\mathcal V}(k,U)]
\,.
\eeq
Projecting this with respect to the test particle 4-velocity leads to
\beq
P(U)k=E(U)\hat {\mathcal V}(k,U)\,,\quad 
U\cdot k=-E(U)
\eeq
so that
\begin{eqnarray}
\fl\quad\label{Frad0}
{\mathcal F}_{\rm (rad)}(U)^\alpha&=&-\sigma \Phi^2 [P(U)^\alpha{}_\beta k^\beta]\, (k_\mu U^\mu)=
\sigma \, [\Phi E(U)]^2\, \hat {\mathcal V}(k,U)^\alpha\,.
\end{eqnarray}
It then follows that the test particle acceleration is aligned with the photon relative velocity in the test particle local rest space,
\beq
a(U)=\tilde \sigma \Phi^2 E(U)^2  \,\hat {\mathcal V}(k,U)\,,
\eeq
where $\tilde \sigma=\sigma/m$. 
Hereafter we
use the simplified notation
$||\nu(U,n)||=\nu$, $\gamma(U,n) =\gamma$, $\hat {\mathcal V}(k,U)=\hat {\mathcal V}$.

From Eq.~(\ref{diff_obg}), after scalar multiplication by $U$ and using Eqs.~(\ref{eq:phot}) and (\ref{polarnu}), one finds the relations  
\beq
\label{photon_en}\fl\qquad
E(U)=\gamma E(n)[1-\nu(U,n)\cdot \hat \nu(k,n)]
= \gamma E(n)[1-\nu\cos(\alpha-\beta)]
\,,
\eeq
which leads to the following expression for the photon direction unit 3-vector
\beq\label{Nu}
\hat {\mathcal V}=\left[\frac{E(n)}{E(U)}-\gamma \right]n + \frac{E(n)}{E(U)}\hat \nu(k,n)-\gamma \nu(U,n)\,.
\eeq
Its frame components $\hat {\mathcal V}=\hat {\mathcal V}{}^{\hat t}n+\hat {\mathcal V}{}^{\hat r}e_{\hat r}+\hat {\mathcal V}{}^{\hat \phi}e_{\hat \phi}$ evaluate to
\begin{eqnarray}
\label{eq:hatVu}
\hat {\mathcal V}{}^{\hat t}&=&\gamma \nu \frac{\cos(\alpha-\beta) -\nu}{[1-\nu\cos(\alpha-\beta)]}
 = \nu (\hat{\mathcal V}{}^{\hat r}\sin\alpha
          + \hat{\mathcal V}{}^{\hat \phi}  \cos\alpha)
\,,\nonumber \\
\hat {\mathcal V}{}^{\hat r}&=&\frac{\sin \beta}{\gamma [1-\nu \cos(\alpha-\beta)]} -\gamma \nu  \sin\alpha \,,\nonumber \\
\hat {\mathcal V}{}^{\hat \phi}&=&\frac{\cos \beta}{\gamma [1-\nu \cos(\alpha-\beta)]} -\gamma \nu  \cos\alpha\,,
\end{eqnarray}
where the second equality of the first line is due to the 
orthogonality of the pair $(\hat{\mathcal V},U)$. 

Finally, it is convenient to evaluate the combination $\Phi^2 E(U)^2$ which appears in the radiation force,
\begin{eqnarray}
\fl\qquad
\Phi^2 E(U)^2
&=& \frac{\gamma^2(1+bN^\phi)^2  \Phi_0^2 E^2}{N^3 \sqrt{g_{\theta\theta}} | b \tan \beta|} [1-\nu\cos(\alpha-\beta)]^2\,.
\end{eqnarray}
In the zero angular momentum limit  $b\to0$, $\beta\to\pi/2$, this reduces to \cite{bijanste}
\beq
\fl\qquad
\Phi^2 E(U)^2\to \frac{ \gamma^2 \Phi_0^2 E^2}{N^2 \sqrt{g_{\theta\theta}g_{\phi\phi}}} (1-\nu\sin \alpha)^2\,.
\eeq
This positive factor multiplies the velocity expression (\ref{Nu}) to become the radiation force on the right hand side of Eq.~(\ref{Frad0}). The last term $-\gamma \nu(U,n)$ in (\ref{Nu}) then leads to a drag force term opposing the particle velocity, but with a coefficient $\Phi^2 E(U)^2$ which at large distances decreases like $1/r^2$, reducing its effectiveness in slowing down the test particle.

A straightforward calculation then shows that the frame components of the 4-acceleration $a(U)$ in the equatorial plane and hence the equations of motion of the test particle are given by
\begin{eqnarray}
\label{eq_fundam}
\fl
a(U)^{\hat t}
&=&   \gamma^2 \nu \sin \alpha 
      \left(a(n)^{\hat r}
         +2\nu\cos \alpha\, \theta(n)^{\hat r}{}_{\hat \phi}\right) 
         + \gamma^3\nu \frac{\rmd \nu}{\rmd \tau}
=\tilde\sigma \Phi^2 E(U)^2 \hat {\mathcal V}^{\hat t}\,,\nonumber \\
\fl
a(U)^{\hat r}
&=& \gamma^2 [a(n)^{\hat r}+k_{\rm (lie)}(n)^{\hat r}\,\nu^2 \cos^2\alpha
            +2\nu\cos \alpha\, \theta(n)^{\hat r}{}_{\hat \phi}]\nonumber \\
\fl
&& + \gamma
 \left(\gamma^2 \sin\alpha \frac{\rmd \nu}{\rmd \tau} 
       +\nu \cos \alpha \frac{\rmd \alpha}{\rmd \tau}\right)
=\tilde\sigma \Phi^2 E(U)^2 \hat {\mathcal V}^{\hat r}\,, \\
\fl
a(U)^{\hat \theta}&=& 0\,, \nonumber \\
\fl
a(U)^{\hat \phi}&=& -\gamma^2 \nu^2 \sin \alpha \cos \alpha\, k_{\rm (lie)}(n)^{\hat r} 
   + \gamma\left(
           \gamma^2 \cos \alpha  \frac{\rmd \nu}{\rmd \tau}-\nu\sin \alpha \frac{\rmd \alpha}{\rmd \tau}\right)
=\tilde\sigma \Phi^2 E(U)^2 \hat {\mathcal V}^{\hat \phi}\,.
\nonumber
\end{eqnarray}
The first of these equations is a linear combination 
$a(U)^{\hat t} = \nu [a(U)^{\hat r}\sin\alpha + a(U)^{\hat \phi}  \cos\alpha ]$
 of the remaining ones due to the orthogonality of $a(U)$ and $U$, while the third is identically satisfied. 
Solving the remaining two nontrivial equations for $\rmd\nu/\rmd\tau$ by taking an appropriate combination
one obtains 
\begin{eqnarray}
\label{dnudtau1}
\fl
\frac{\rmd \nu}{\rmd \tau}
&=& 
-\frac{\sin\alpha}{\gamma}[a(n)^{\hat r}+2\nu\cos \alpha\, \theta(n)^{\hat r}{}_{\hat \phi}]
   \nonumber \\
\fl
&& + \frac{A(1+bN^\phi)}{N^2(g_{\theta\theta} g_{\phi\phi} )^{1/2} | \sin \beta|} 
[\cos(\alpha-\beta) -\nu][1-\nu\cos(\alpha-\beta)] \,,
\end{eqnarray}
where
$A = \tilde\sigma \Phi_0^2 E^2 = (\sigma/m) \Phi_0^2 E^2 $ is a positive constant for a given fixed radiation field.
For a black hole spacetime where $\sqrt{g_{\phi\phi}}=r$, the luminosity at infinity is $L_\infty=4\pi E^2 \Phi_0^2$, while the Eddington luminosity at infinity is $L_{\rm Edd}= 4\pi Mm/\sigma$ \cite{abr-ell-lan}, this constant satisfies
\beq
  \frac{A}{M} = \frac{L_\infty}{L_{\rm Edd}} \,.
\eeq

Backsubstituting this solution into the radial equation and solving for $\rmd\alpha/\rmd\tau$ leads to
\begin{eqnarray}\label{dalphadtau1}
\frac{\rmd \alpha}{\rmd \tau}
&=& -\frac{\gamma\cos \alpha}{\nu}[a(n)^{\hat r} +2\nu\cos \alpha\, \theta(n)^{\hat r}{}_{\hat \phi} +\nu^2 k_{\rm (lie)}(n)^{\hat r}] 
\nonumber \\
&&
+\frac{A}{\nu }  
\frac{  (1+bN^\phi) [1-\nu\cos(\alpha-\beta)] }{N^2(g_{\theta\theta} g_{\phi\phi} )^{1/2} |\sin \beta|}  \sin(\beta -\alpha)\,.
\end{eqnarray}
In the zero angular momentum limit $b\to 0$, $\cos\beta\to0$, then 
$\sin\beta\to\pm1$,
$\cos(\alpha-\beta)\to \pm\sin\alpha$,
$\sin(\beta-\alpha)\to \pm\cos\alpha$ and
these two equations reduce to those in \cite{bijanste}.

Finally, we have in addition the remaining two equations
\beq
\frac{\rmd r}{\rmd \tau}=
\frac{\gamma \nu\sin\alpha}{\sqrt{g_{rr}}}\,,
\qquad
\frac{\rmd \phi}{\rmd \tau}=
\frac{\gamma}{\sqrt{g_{\phi\phi}}} (\nu\cos\alpha -\nu_{(s)})
\,.
\eeq

This system of four differential equations for $\nu$, $\alpha$, $r$ and $\phi$ admits
a critical solution at a  radial equilibrium which corresponds to a circular orbit of constant radius $r=r_0$, constant speed $\nu=\nu_0$, and constant angles $\beta=\beta_0$, $\alpha=\alpha_0$.
The constancy of the radius requires  $\sin\alpha_0=0$, $\cos\alpha_0=\pm1$  and therefore
$\sin(\beta_0-\alpha_0)=\cos\alpha_0 \sin\beta_0$ and
$\cos(\alpha_0-\beta_0)=\cos\alpha_0 \cos\beta_0$.
For circular orbits it is convenient to reintroduce the azimuthal velocity component $\nu^{\hat\phi}_0=\nu_0\cos\alpha_0=\pm \nu_0$. This occurs in the conditions $\rmd \nu/\rmd\tau =0 =\rmd \alpha/\rmd\tau$ for
such a critical orbit to exist, which due to Eqs.~(\ref{dnudtau1}) and (\ref{dalphadtau1}) imply respectively
\begin{eqnarray}
\label{dnudtau}\fl\qquad
0&=& \frac{A(1+bN^\phi)}{N^2(g_{\theta\theta} )^{1/2}| \sin \beta_0| } 
(\cos\beta_0 -\nu^{\hat\phi}_0)(1 -\nu^{\hat\phi}_0\cos \beta_0 ) \,,\\ \fl\qquad
0&=& \frac{\gamma_0 }{\nu_0}
        \left[ -[a(n)^{\hat r} +2\nu^{\hat\phi}_0\, \theta(n)^{\hat r}{}_{\hat \phi} 
                        +(\nu^{\hat\phi}_0)^2 k_{\rm (lie)}(n)^{\hat r}] \right.
\nonumber \\ 
\label{dalphadtau}\fl\qquad
&&\left.
+  
\frac{ A\,{\rm sgn}(\sin \beta_0) (1+bN^\phi) \gamma_0^{-1} (1 -\nu^{\hat\phi}_0\cos \beta_0 )}
      {N^2(g_{\theta\theta} g_{\phi\phi})^{1/2}}    \right]
\,,
\end{eqnarray}
where it is understood that all functions of $r$ are evaluated at $r=r_0$.
The first  Eq.~(\ref{dnudtau}) simply equates the particle azimuthal velocity to the photon azimuthal velocity,
\beq
\label{nu0limit}
\nu^{\hat\phi}_0 = \cos \beta_0 
       = \frac{ b N}{\sqrt{g_{\phi\phi}}(1+bN^\phi) }
\rightarrow 
  \gamma_0= 1/|\sin\beta_0|
\,,
\eeq
so that the circular orbit particle impact parameter (\ref{bppcirc}) coincides with the photon impact parameter (\ref{bp}).
This makes the photon relative velocity with respect to the test particle (\ref{eq:hatVu}) purely radial, reducing to 
$( \hat {\mathcal V}{}^{\hat t},\hat {\mathcal V}{}^{\hat r},\hat {\mathcal V}{}^{\hat \phi})= (0,{\rm sgn}(\sin\beta_0),0)$.
In fact one could start from this obvious requirement on the photon direction in the particle rest space for a circular orbit at constant speed to be consistent, and insert these simple values into the acceleration Eq.~(\ref{eq_fundam}) to directly get the conditions that such an orbit exist.

The remaining differential equation (\ref{dalphadtau}) for the radial acceleration implies
\begin{eqnarray}\label{forcebalance}
\fl\qquad
a(n)^{\hat r} + 2\nu^{\hat\phi}_0\, \theta(n)^{\hat r}{}_{\hat \phi} + \,  k_{\rm (lie)}(n)^{\hat r} \nu^{\hat\phi}_0{}^2
=
 \frac{  A\,{\rm sgn}(\sin \beta_0)(1+bN^\phi) \gamma_0^{-3} }{N^2(g_{\theta\theta}g_{\phi\phi} )^{1/2} }   
\,. 
\end{eqnarray}
Introducing the corotating and counterrotating circular geodesic azimuthal velocities $\nu_+$ and $\nu_-$ and re-expressing the $b$ factor in terms of the speed through the relation
$1+bN^\phi=(1-   \nu^{\hat\phi}_0{}/\bar\nu_{\rm(s)})^{-1}>0$, 
we have
\begin{eqnarray}\label{forcebalance2}
&&
a(n)^{\hat r}\left( 1-\frac{\nu^{\hat\phi}_0{}}{\nu_+}\right)\left( 1-\frac{\nu^{\hat\phi}_0{}}{\nu_-}\right)
= \frac{  A\,{\rm sgn}(\sin \beta_0) \gamma_0^{-3} }
      { N^2(g_{\theta\theta}g_{\phi\phi} )^{1/2} (1-  \nu^{\hat\phi}_0{}/\bar\nu_{\rm(s)})}  
\,. 
\end{eqnarray}
This relation gives the critical circular velocity $\nu^{\hat\phi}_0{} $ implicitly as a function of the critical radius $r_0$, but only when one replaces it in  terms of $b$ using (\ref{nu0limit}) does one get an implicit relationship determining the critical radius $r_0$ itself as a function of $b$.

Solving (\ref{forcebalance2}) for the ratio $A/M$ one finds
\begin{eqnarray}\label{forcebalance3}
\fl
&&
{\rm sgn}(\sin \beta_0)\frac{A}{M} 
= N^2(g_{\theta\theta}g_{\phi\phi} )^{1/2} \frac{a(n)^{\hat r}}{M}  
        \gamma_0^{3} \left( 1-\frac{\nu^{\hat\phi}_0{}}{\nu_+}\right)\left( 1-\frac{\nu^{\hat\phi}_0{}}{\nu_-}\right)  
                        \left(1-  \frac{\nu^{\hat\phi}_0{}}{\bar\nu_{\rm(s)}}\right) 
\,.
\end{eqnarray}
This formula has zeros at the two geodesic velocities, while the last factor with $\bar{\nu}_{\rm (s)}$ is restricted to positive values, corresponding to $\nu^{\hat\phi}_0>\bar{\nu}_{\rm (s)}$.

The terms in Eq.~(\ref{forcebalance}) from left to right respectively (once all moved to the right hand side) may be interpreted as a radial force (per unit mass) balance relation between the inward gravitational force, a gravitomagnetic force, the outward centrifugal force, and the outward radiation force \cite{idcf2} as seen by the ZAMOs. Consider the outgoing photon case $\sin\beta>0$. Increasing the radiation force from zero (where the equations of motion describe circular geodesic motion), 
the additional positive radial radiation force allows circular orbits at a given radius (greater than that of the photon circular orbits) to have a smaller centrifugal force term and hence smaller speed than in the geodesic case. 
In black hole spacetimes this turns out also to allow critical circular orbits to occur at smaller radii than in the geodesic case, where timelike geodesics exist only outside the radius at which the first null circular geodesic orbit occurs as one decreases the radius.

\begin{table}
\begin{center}
Table 1. Metric and ZAMO kinematical quantity expressions for the equatorial plane of the Kerr spacetime, where $\Delta=r^2-2Mr+a^2$.\\{}
\end{center}
\begin{center}
\begin{tabular}{|l|c|} 
\hline\Strut 
ZAMO quantity & Kerr     \\
\hline\Strut 
$N=(-g^{tt})^{-1/2}$ & $[r\Delta /(r^3+a^2r+2a^2M)]^{1/2}$   \\
\hline\Strut   
$N^\phi=N_\phi/g_{\phi\phi}$ & $-2aM/(r^3+a^2r+2a^2M)$ \\
\hline\Strut  
$N_\phi=g_{t\phi}$ & $-2aM/r$   \\
\hline\Strut  
$g_{rr}$ & $r^2/\Delta$   \\
\hline\Strut  
$g_{\phi\phi}$ & $(r^3+a^2r+2a^2M)/r$  \\
\hline\Strut  
$a(n)^{\hat r}$&  $M[(r^2+a^2)^2-4a^2Mr]/[r^2 \Delta^{1/2}(r^3+a^2r+2a^2M)]$   \\
\hline\Strut  
$\theta(n)^{\hat r}{}_{\hat\phi}$& $-aM(3r^2+a^2)/[r^2(r^3+a^2r+2a^2M)]$\\
\hline\Strut  
$k_{(\rm lie)}(n)_{\hat r}$ & $- \Delta^{1/2}(r^3-a^2M)/[r^2(r^3+a^2r+2a^2M)]$\\
\hline\SStrut  
$\nu_{\rm(s)}$, $\nu_\pm$ & $-\frac{2aM}{r\Delta}$, $\frac{r^2+a^2\mp 2a \sqrt{Mr}}{\Delta^{1/2} (a\pm r\sqrt{r/M})} $\\
\hline 
\end{tabular}
\end{center}
\end{table}

Note that in the limit $A\to0$ so that the particle moves along a circular geodesic, the solution (\ref{nu0limit}) of the constant speed equation (\ref{forcebalance}) is no longer relevant since that equation is identically zero, while the solution of the radial force balance equation (\ref{forcebalance}) 
 in terms of $\nu_0$ leads to the Keplerian circular orbit azimuthal velocity formula for $\nu^{\hat\phi}_0{}=\nu_\pm$ as a function of $r_0$. 
However, if left expressed in terms of the photon impact parameter $b$, which coincides with the particle impact parameter $b_{\rm(p)}$ when (\ref{nu0limit}) holds, one gets the circular orbit particle impact parameter versus radius relation. In other words,
as one increases $A$ from 0, the  geodesic circular orbit relation of particle impact parameter versus radius is continuously deformed to describe the photon impact parameter versus the critical circular orbit radius. The previous discussion then explains how this deformation takes place, as will be seen explicitly for the exterior field of black hole spacetimes.

Note that for an outgoing flux ($\sin\beta_0>0$), $A>0$ requires the azimuthal critical velocity to be confined to the interval  $\nu_- < \nu^{\hat\phi}_0{} <\nu_+$ between the two geodesic velocities. 
As one increases the radius in a black hole spacetime, the remaining factors in the expression (\ref{forcebalance3}) for $A/M$ go to 1, which leads to $A/M\approx1$ for a critical circular orbit to exist at large radii. 
It will turn out that $A/M<1$ is necessary for radii not too close to the horizon. Of course in a physical application, the photon emitting surface must lie somewhere outside the horizon, so a cutoff radius $R$ would have to be introduced, like the radius $R/M=3$ for timelike circular photon orbits in the Schwarzschild case \cite{abr-ell-lan}, where azimuthally emitted photons are trapped in circular orbits.

\begin{figure} 
\begin{center}
\includegraphics[height=2.5in,width=3.5in]{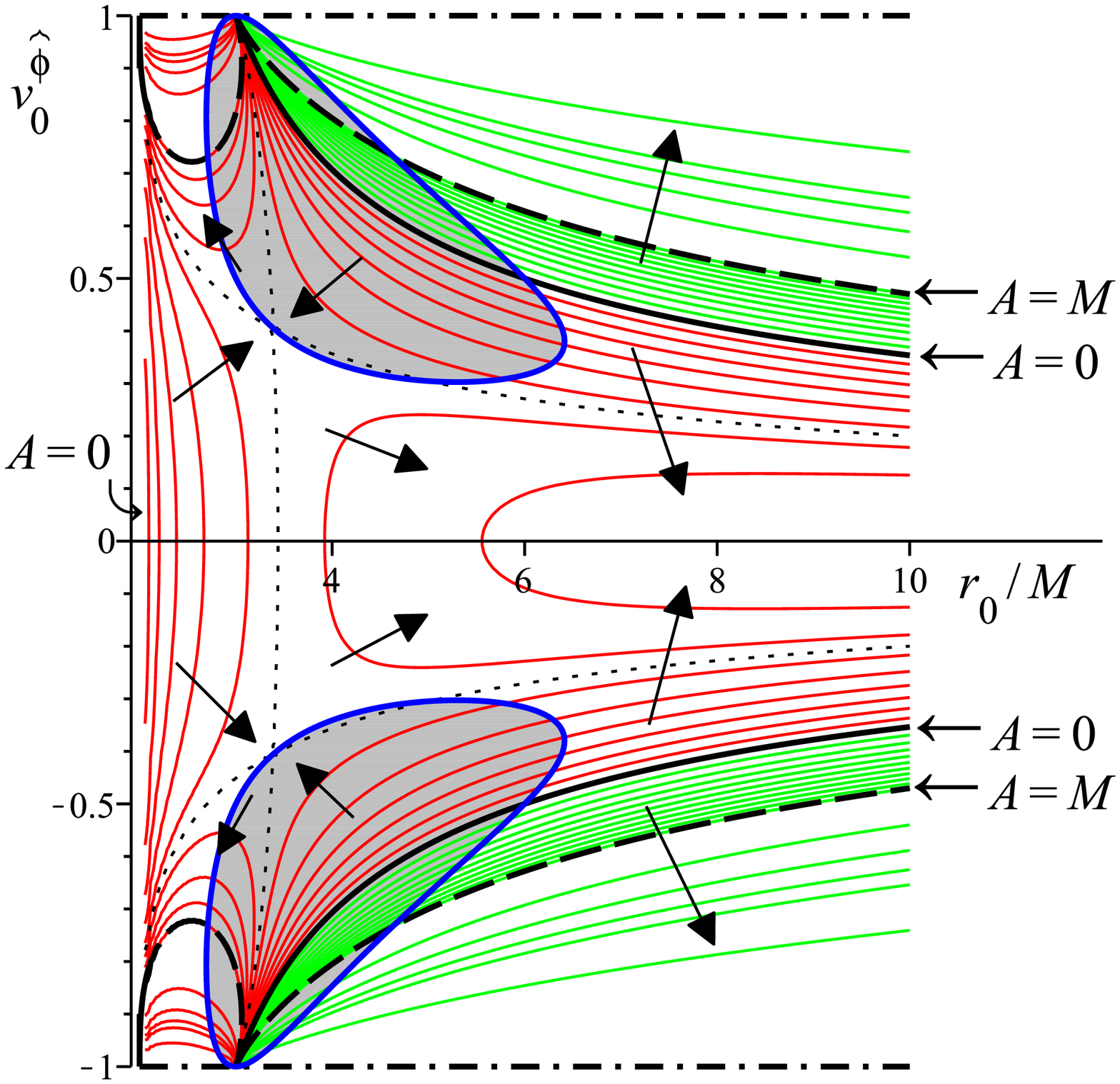}
\includegraphics[height=2.5in,width=3.5in]{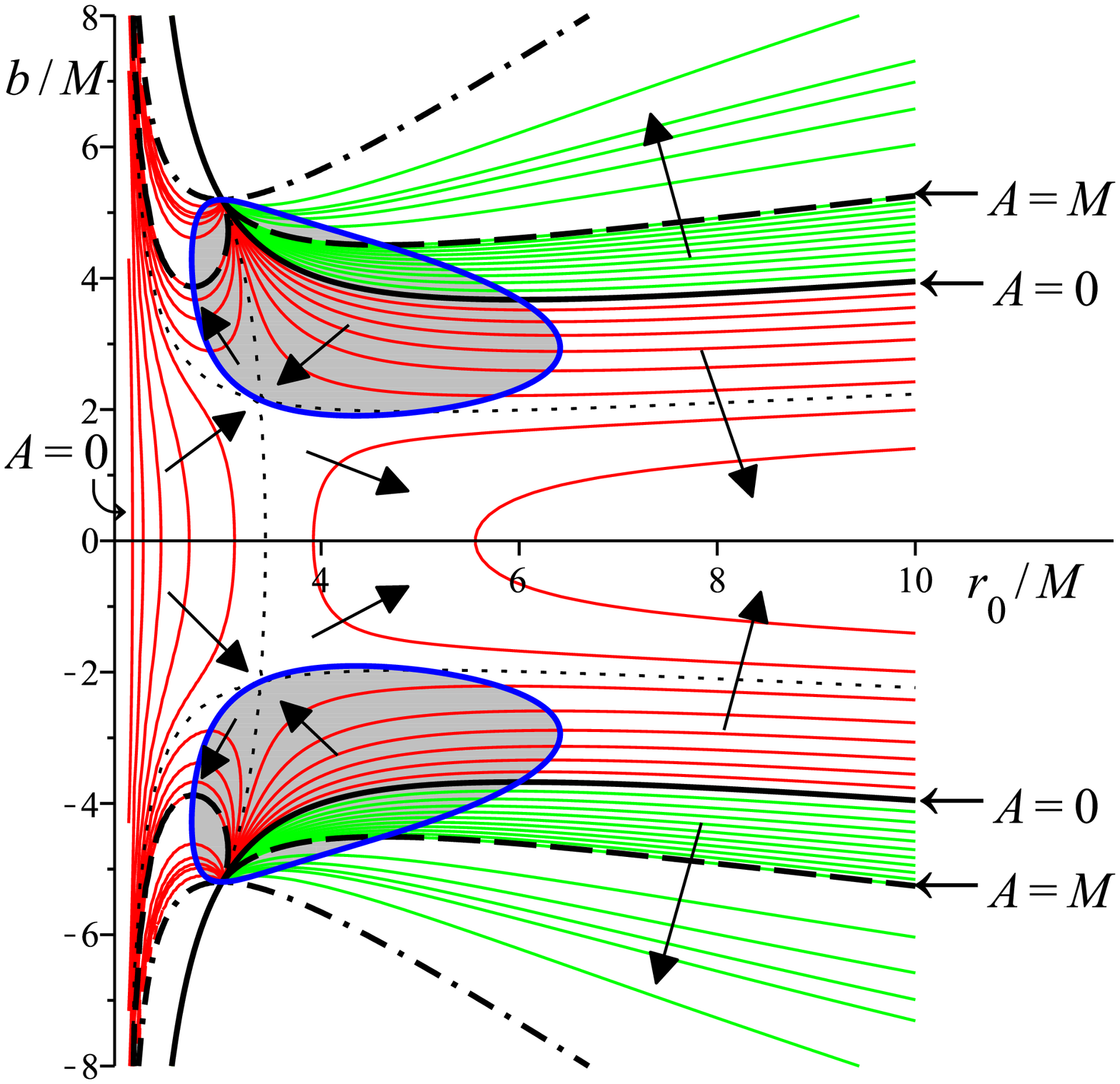}
\end{center}
\caption{
Top:
The critical azimuthal velocity $\nu^{\hat\phi}_0{}$ versus critical radius $r_0/M> 2$ for the Schwarzschild case for selected values of $A/M$. Physical velocities are confined to the interval $-1< \nu^{\hat\phi}_0{}<1$, with $A/M\to\infty$ corresponding to $|\nu^{\hat\phi}_0{}|=1$ indicated by the thick dot-dash lines, and the thick solid curves indicating the geodesic velocities corresponding to $A=0$, enclosing the outgoing photon region, outside of which is the ingoing photon region.  
The thick closed loop curves enclosing the shaded regions are explained below. 
For context the thick dashed curves correspond to $A/M=1$. Arrows indicate the direction of increasing values of $A$. Note the two pairs of accumulation points of the family of curves near the horizon at unit velocity.

Bottom: 
The same figure with the vertical axis transformed by relation (\ref{nu0limit}) instead to show the particle impact parameter $b/M$, with the accumulation points at the horizon pushed out to infinity.
The thick closed loop curves enclosing the shaded regions join together points corresponding to the critical points of the constant $A$ curves of $b$ versus $r_0$. For each curve of a given $A$ value, its intersection with the horizontal line of a given $b$ value locates the radii $r_0$ at which the critical orbits occur. Those horizontal curves which pass through the interior of the closed loops correspond to the case in which three critical radii exist, with the unstable orbit of the three lying in the interior, while those passing outside these loops correspond to the single stable critical radius case.
 } 
\label{fig:radiusnu}
\end{figure}

\section{Critical orbit radius and velocity relations for black holes} 

The general relations obtained in the previous section specialize to the Kerr case and then further to the Schwarzschild case using the ZAMO kinematical quantity expressions listed in Table 1. 
The outer horizon radius and the ergosphere radius in the equatorial plane of the Kerr spacetime are respectively
$ r_{\rm(h)}= M+\sqrt{M^2-a^2}$
and $ r_{\rm(erg)}= 2M$. We only examine the region outside the outer horizon.

The explicit relationship between the orbital velocity $\nu^{\hat\phi}_0$ of the critical circular orbits and the radius $r_0$ is simpler in the Schwarzschild case where the corotating ($\alpha_0=0$) and counterrotating ($\alpha_0=\pi$)  orbits have the same speed $\nu_0$. The rotation of the Kerr metric breaks this symmetry and complicates the relationship, so it is useful to start with the former case.
The force balance equation for the critical circular orbits is then explicitly (with $\nu_-=-\nu_+ $)
\begin{eqnarray}\label{forcebalance2s}
 N\gamma_0^{3}\left( 1-\frac{\nu_0^2}{\nu_\pm^2}\right)
&=&
 {\rm sgn}(\sin \beta_0)\frac{A}{ M }  
\,. 
\end{eqnarray}
Note that as one approaches the horizon, $N\rightarrow 0$ and $|\nu_\pm|\rightarrow\infty$, so to compensate one must have $\gamma_0\rightarrow\infty$ and thus $\nu_0\rightarrow 1$ for any nonzero value of $A$. Hence, $(r_0/M,\nu_0^{\hat\phi}) =(2,\pm 1)$ are accumulation points for the curves of constant $A$, because the curves for all possible values of $A$ converge there. Similarly, at the null circular geodesic radius $r_0/M=3$ where $|\nu_\pm|\rightarrow 1$, one has another such null accumulation point, because $\gamma_0\rightarrow\infty$ for $\nu_0\rightarrow 1$ irrespectively of the value of $A$.

Fig.~\ref{fig:radiusnu} shows the family of curves describing the critical circular orbit parameters $\nu^{\hat\phi}_0$ and $b/M$ versus radius $r_0/M$ for equally spaced values of $A$ from 0 to 1 at intervals of $0.1$ and thereafter values of 1, 2, 3, 4, 5, 10, starting at the horizon $r_0/M=2$ in horizontal units of $r_0/M$.
Just from the overall sign of both sides of Eq.~(\ref{forcebalance2s}), one sees that for the outgoing/ingoing photon cases $\sin\beta_0>0,\sin\beta_0<0$, the critical speed is less than/greater than the geodesic speed.
The two thick solid curves are the geodesic velocities corresponding to $A=0$, which confine between them all the curves of constant $A$ for the outgoing photon case, while the outgoing case curves lie outside this region.
The curves of constant $A$
start at the value 0 at the horizon (vertical axis) and on the geodesic velocity curves and increase in value in the directions indicated by the arrows, meeting at the saddle point
on the separatrix curve at $A/M\approx0.647$ (shown as a set of three thin small dashed component curves, two relatively horizontal, one relatively vertical, dividing the region between the geodesic curves into 6 sectors), and continuing to increase in value past that point, with the limit $A/M\to1 $ occurring at $r\to\infty$, while $A/M $ increases without limit in the inner sectors near the horizon.
For context the thick dashed curve corresponds to $A/M=1$, a value which is never reached in between the geodesic curves to the right of the vertical component of the separatrix curve.
For outgoing photons as one increases $A/M$ from 0 towards 1, the curves of constant $A/M$ move farther out to larger radii, moving out to infinity in the limit $A/M\to1$, corresponding to the fact that for $A/M>1$, test particles are pushed out to infinity in this region.
Since $\nu^{\hat\phi}_0{}=bN/r$ for the Schwarzschild case, the limiting null accumulation points $\nu^{\hat\phi}_0{}=\pm1$ at the horizon ($N\rightarrow 0$) correspond to $b\rightarrow\pm\infty$.

In terms of the photon impact parameter the force balance condition is the following 
\beq\label{AMN}\fl\qquad
\frac{A}{M N}
=\frac{1-(r_0/M) N^2 \cos^2\beta_0}{\sin^3\beta_0}
={\rm sgn}(\sin\beta_0) \displaystyle \frac{1-\displaystyle\frac{ b^2}{M r_0 }\left(1-\displaystyle\frac{2M}{r_0}\right)^2 }
{\left[ 1-\displaystyle\frac{b^2}{r_0^2}\left(1-\displaystyle\frac{2M}{r_0}\right) \right]^{3/2}}
   \,.
\eeq
This last equation determines the critical radius in terms of the photon impact parameter and then through (\ref{nu0limit}) in terms of the critical  angular speed.
For fixed values of $b$ and $A$, one or more values of $r_0$ may satisfy it. 
For the curve of fixed values of both $A$ and sgn($\sin\beta_0$) in the bottom plot of Fig.~\ref{fig:radiusnu}, the number of its intersections with the horizontal line at each value of $b$ indicates the number of critical radii which exist for that case. The thick loop curves enclose the region of unstable intermediate radius orbits for those values of $b$ for which three critical radii exist. Appendix A discusses the stability of the critical orbits. 

In the lower diagram of Fig.~\ref{fig:radiusnu}, as one increases the value of $|b|$ from 0 where the constant $A$ curves have no critical points (at which $db/dr_0=0$) and are cut by a single horizontal line, at a certain point when the loop curve is encountered, such a critical point develops (leading to one additional critical radius) and bifurcates into a pair of local extrema (one to the left, one to the right but for distinct values of $A$ as one moves along the loop curve away from the horizontal axis); the local extremum is then cut an additional two times by each horizontal line for a total of three distinct radii at which critical orbits exist for a given value of $A$. These additional extrema disappear when $b$ reaches the value corresponding to the $r_0=3M$ accumulation point at the farthest point on the loop curve from the horizontal axis.

Naively speaking using the space plus time language of spatial forces and only considering the outgoing photon case, the net outward radial force profile as a function of $r$  decreases from positive to negative values through a single stable equilibrium radius where it vanishes (balancing the outward force of the outgoing radiation with the inward gravitational force), so that at nearby radii the radial force pushes the test particle towards the equilibrium radius. As one changes the parameters of the problem, when this radial force profile function drops to the horizontal axis and crosses to create two more equilibrium radii, the intermediate radius has the opposite behavior of the sign of the net force as one increases the radius through it, pushing the test particle away from the equilibrium point, so it must be unstable. This is confirmed by the detailed stability analysis of Appendix A which shows that when three such equilibrium radii exist, the intermediate value corresponds to an unstable orbit, while the outer ones correspond to stable orbits.


\begin{figure} 
\begin{center}
\includegraphics[height=2.5in,width=3.5in]{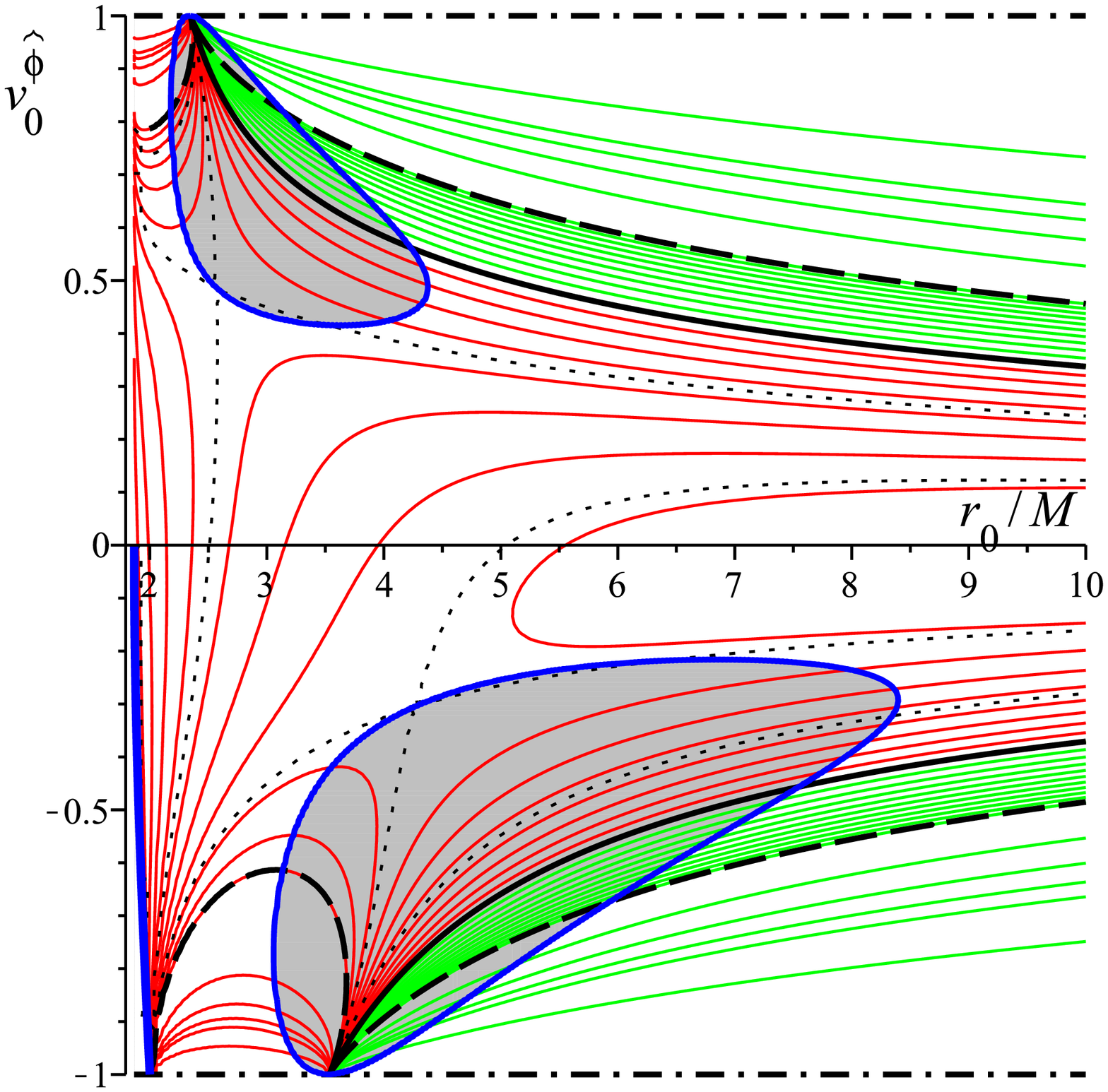}
\includegraphics[height=2.5in,width=3.5in]{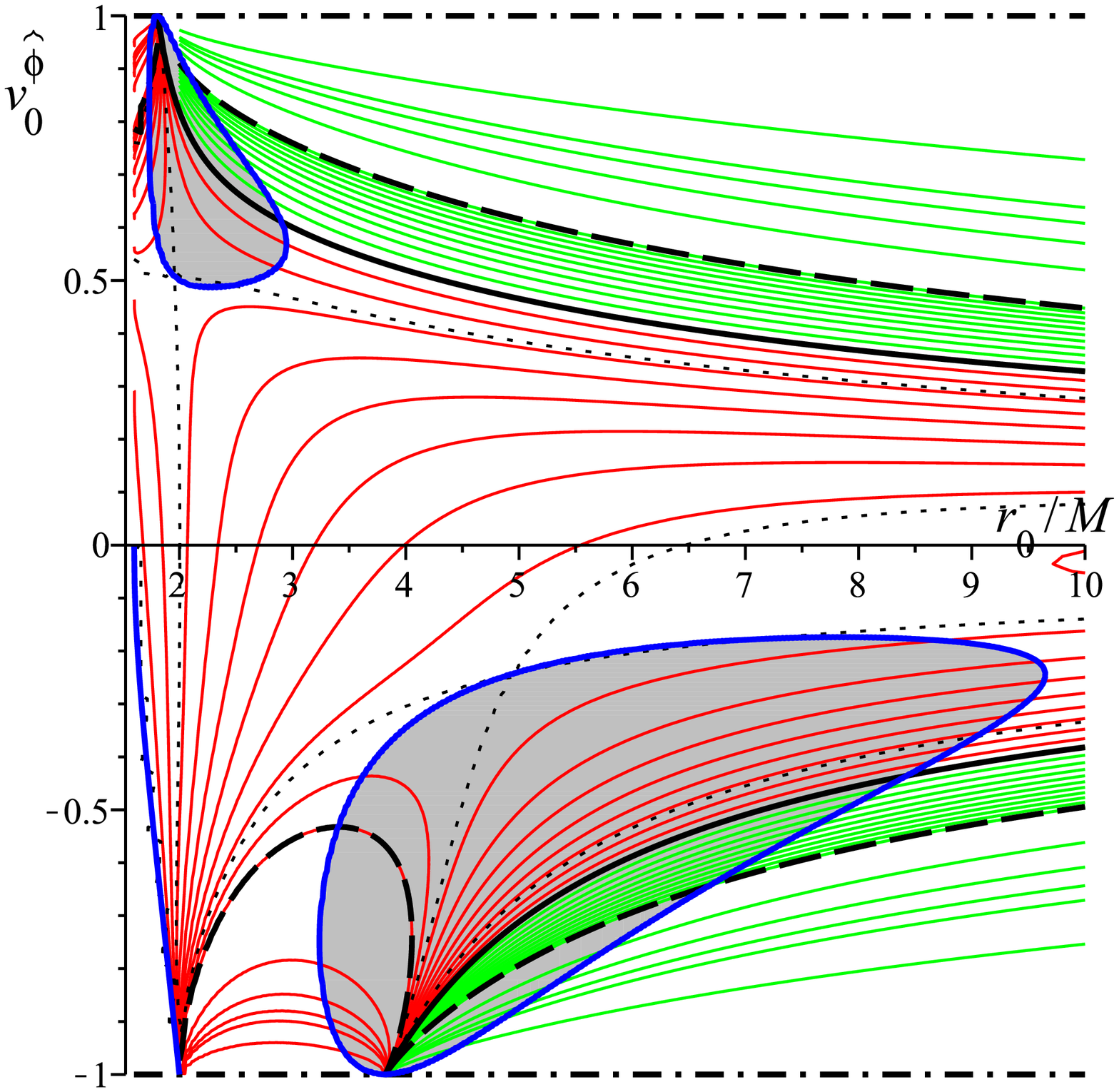}
\includegraphics[height=2.5in,width=3.5in]{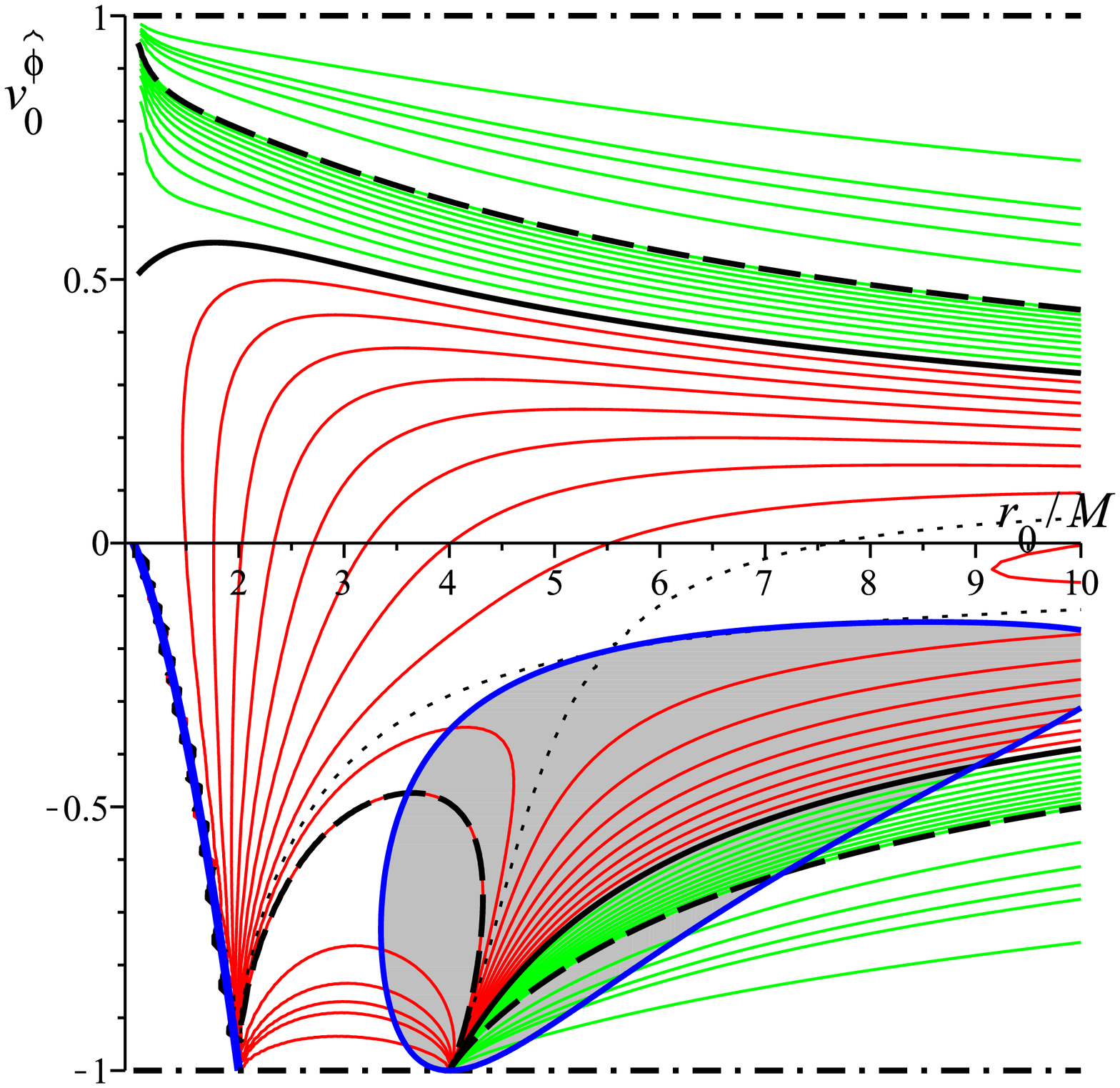}
\end{center}
\caption{
The critical azimuthal velocity $\nu^{\hat\phi}_0{}$ versus critical radius $r_0/M$ for the Kerr case for equally spaced values of $A$, with the same conventions as in Fig.~\ref{fig:radiusnu}, for $r_0>r_{\rm(h)}$. 
Top: the case $a/M=0.5$.
Middle: the case $a/M=0.8$. 
Bottom: the extreme case $a/M=1$.
} 
\label{fig:radiusnukerr1}
\end{figure}

\begin{figure} 
\begin{center}
\includegraphics[height=2.5in,width=3.5in]{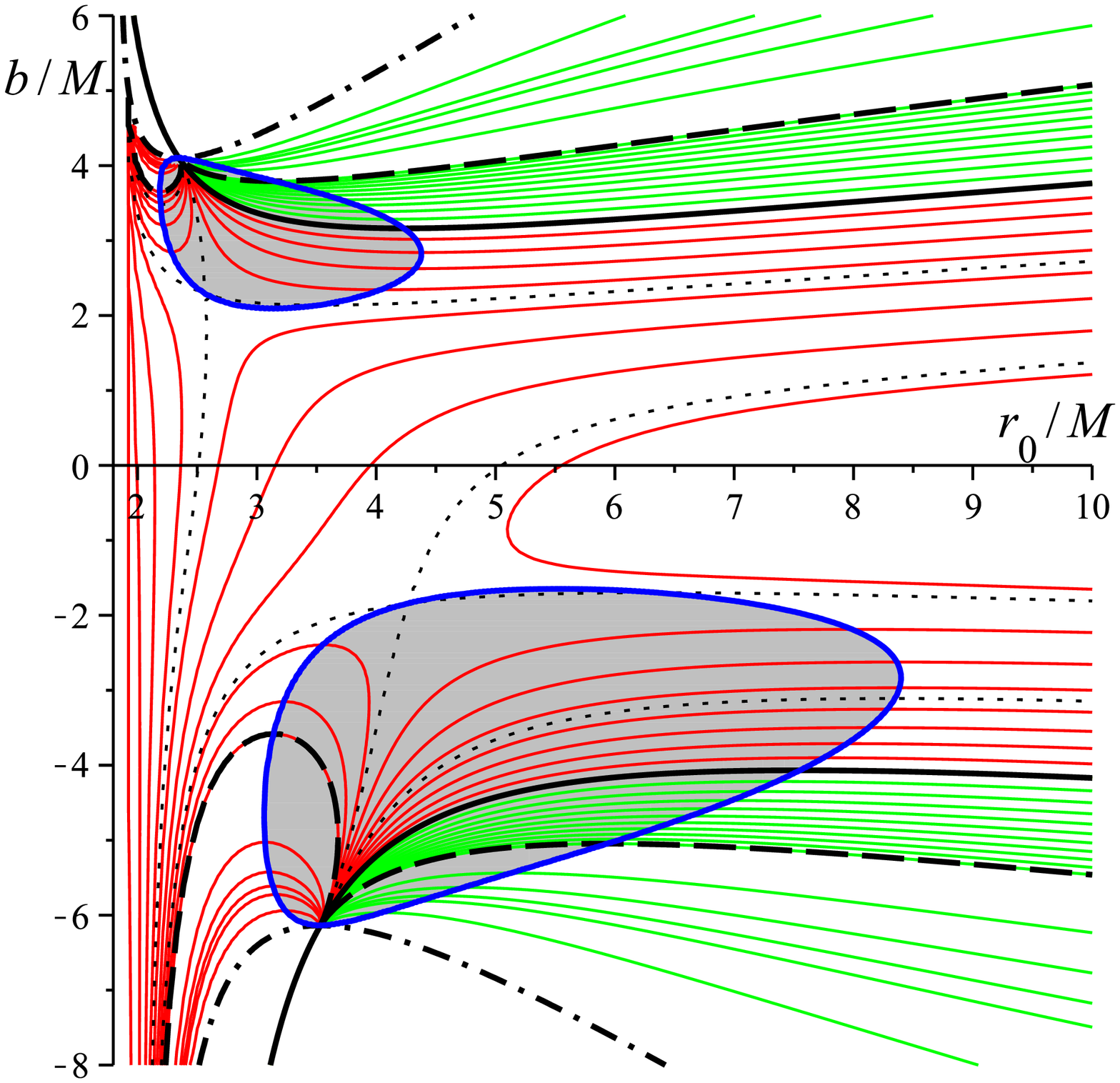}
\includegraphics[height=2.5in,width=3.5in]{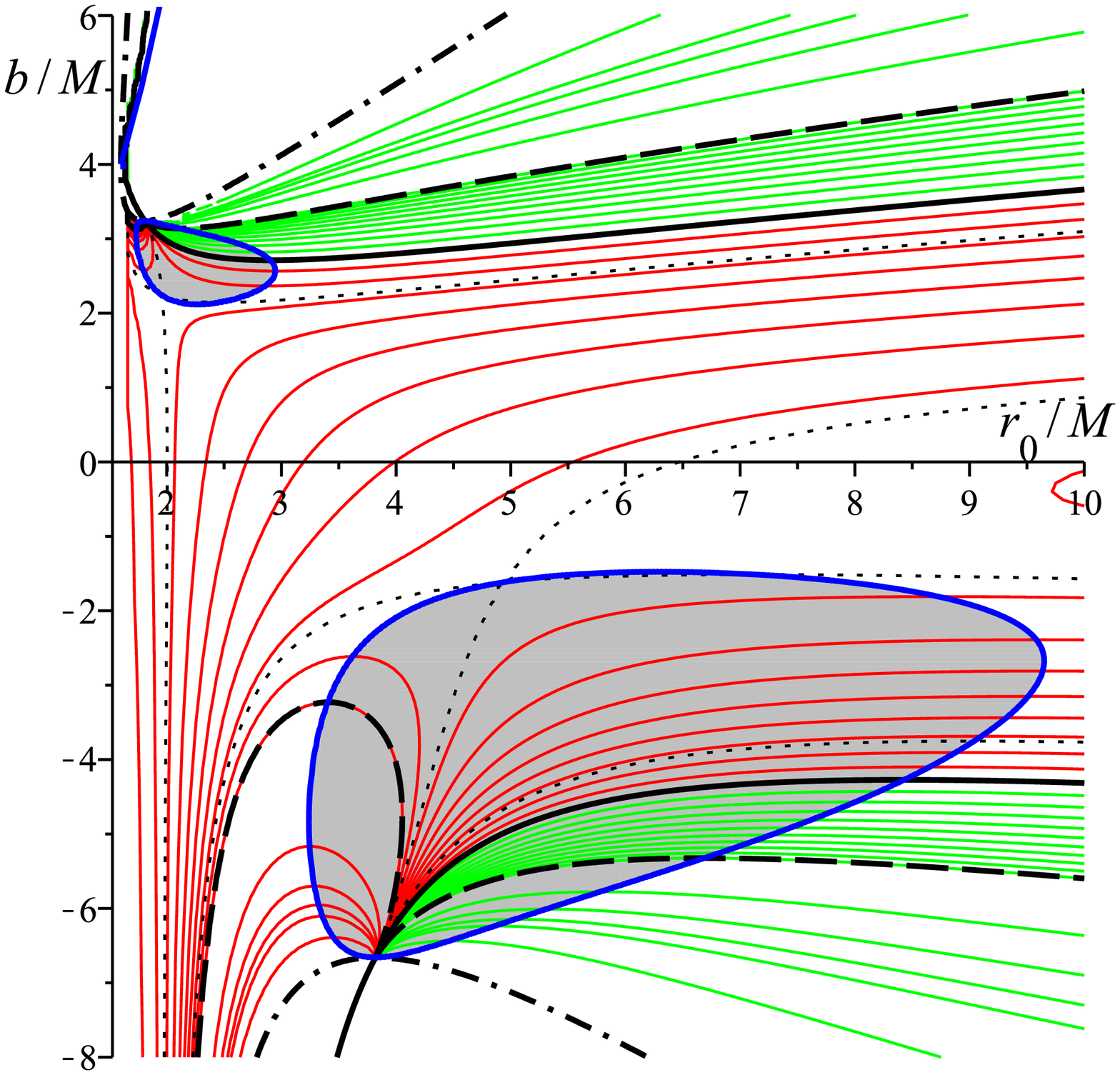}
\includegraphics[height=2.5in,width=3.5in]{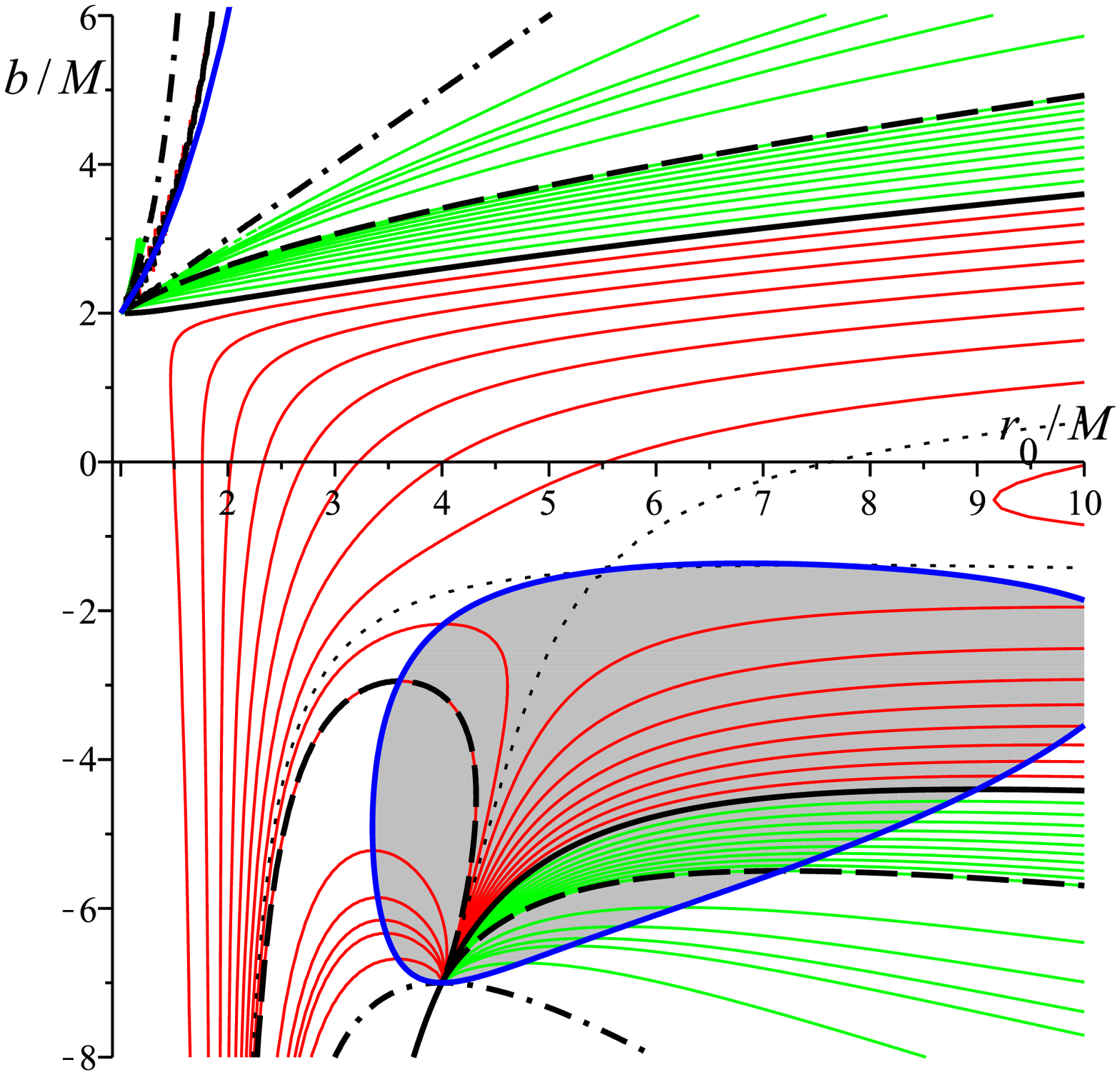}\end{center}
\caption{

The Kerr case showing the photon impact parameter $b/M$  versus radius $r_0/M$ instead of the azimuthal velocity, to be compared with the $a=0$ case in Fig.~ \ref{fig:radiusnu}, again for $r_0>r_{\rm(h)}$.
Top: the case $a/M=0.5$. 
Middle: the case $a/M=0.8$.
Bottom: the extreme case $a/M=1$.
} 
\label{fig:radiusbkerr1}
\end{figure}


In the case $b=0$ and $\sin\beta_0>0$ of purely radial outward photon motion, Eq.~(\ref{AMN}) reduces to the result 
\beq
\frac{A}{M} = N = \left(1-\frac{2M}{r_0}\right)^{1/2}
\eeq 
of Bini et al \cite{bijanste},
which requires  $A/M<1$ for a solution to exist. This is the simple situation near the horizontal axis in Fig.~\ref{fig:radiusnu} where increasing $A/M$ from 0 to 1  moves the single critical radius from the horizon out to infinity. 
For $|b/M|$ less than about 2 this situation persists, but as $|b/M|$ reaches a value near 2 
(just before reaching the dotted separatrix crossing point at $A/M\approx0.647$ near $r_0/M =3.4$), the number of critical radii for a given value of $A$ jumps from 1 to 3, making the qualitative behavior of the orbits much more complicated. Two radii occur inside (left of) the vertical component of the separatrix curve, while the third moves out to larger values of $r_0$ as $A$ increases in the zone between the geodesic curves. 
Numerical sampling of initial data indicates that the innermost and outermost radii of the three represent (locally) stable radial equilibrium points, but the intermediate critical orbit is unstable, and the innermost critical orbit wins out if the particle does not have suitable initial data to be injected into the outer critical orbit. 
Thus some particles will fall into the outer critical orbit and others into the inner critical orbit, or in the case of a cutoff radius $R$ larger than the critical orbit, into the emitting surface. 
Appendix A briefly discusses the stability of the critical orbits, confirming the numerical results.

Figs.~\ref{fig:radiusnukerr1} and \ref{fig:radiusbkerr1} show how turning on the Kerr rotation parameter $a$ from 0 to the extreme value $M$ distorts the Schwarzschild case diagram.
This parameter introduces an asymmetry between positive and negative azimuthal velocities, clearly shown in the deformation of the shaded regions of unstable intermediate critical orbits, with the comoving such region disappearing into the outer horizon as the extreme value $a=M$ is reached.
As $a/M$ is increased from 0 to 1, the outer radius of the ergosphere remains at $r/M=2$ while the outer horizon moves from that radius to  $r/M=1$, while the upper corotating null geodesic accumulation point moves from $r/M=3$ to the outer horizon and the lower counterrotating null geodesic accumulation point and the inner counterrotating null accumulation point move away from the horizon.
This latter point is connected to the horizontal axis at the horizon by  the new thick solid curve
representing the zero of the equilibrium condition factor at $\bar\nu_{\rm(s)}$, on and below which there are no relevant equilibrium solutions.

\begin{figure} 
\begin{center}
\includegraphics[scale=0.4]{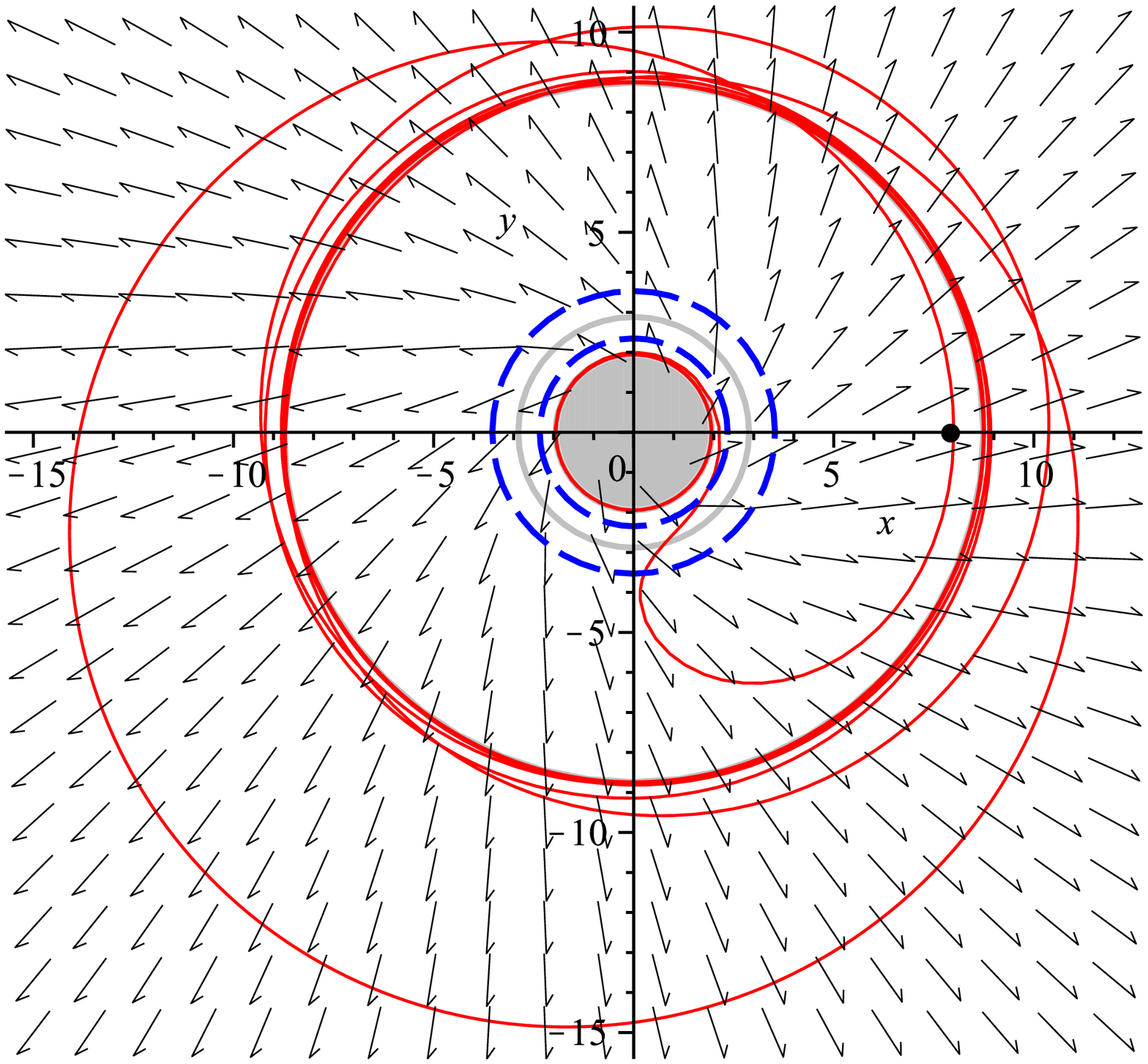}
\end{center}
\caption{
The spacetime and photon parameters are  $a/M=0.5$, $A/M=0.3$, $b/M=3$, showing two orbits moving initially from the bullet point on the horizontal axis in the two azimuthal directions just inside the outer critical radius with 1.2 times the critical speed for that counterclockwise critical orbit. 
The unit velocity direction field $\hat\nu(k,n)$
of the radiation with respect to the ZAMOs is superimposed on the plot, showing the additional counterclockwise rotation of the photon trajectories with respect to the counterclockwise rotating ZAMOs. 
The dashed circles are the two null circular geodesics orbits. The gray filled circle extends to the horizon. 
The counterclockwise moving orbit settles down to the outermost critical orbit, while the clockwise moving orbit quickly falls into the innermost critical orbit near the horizon. The gray circle between the null orbits is the unstable critical orbit.
This axes show units of $r/M$.
} 
\label{fig:four}
\end{figure}

\section{Orbits for Kerr black holes} 

For small enough values of $b$ where there is only one unique critical radius, the situation is not much different from the $b=0$ case qualitatively, apart from the nonzero critical velocity equal to the photon azimuthal velocity at the critical radius. However, as one increases $b$ there comes a point where there are 3 distinct critical radii, apparent from the intersection of horizontal lines with the curves of a given $A$ in Fig.~\ref{fig:radiusbkerr1}.

Fig.~\ref{fig:four} shows a pair of oppositely directed particles initially on the positive horizontal axis at the bullet point just inside the outermost (counterclockwise) critical orbit in a $a/M=0.5$ Kerr black hole with photon parameters $b/M=3, A/M=0.3$, with a bit more than the critical azimuthal speed. The initially clockwise directed particle settles into the counterclockwise critical orbit, but the initially clockwise directed particle quickly drops into the innermost critical orbit. If one has a cutoff radius $R$ for the photon emission surface which is larger than the innermost critical orbit radius, this means some particles will drop into that surface while others will find their way into the outer critical orbit. The outer critical orbit appears to be only locally stable in the sense that a test particle at that critical radius must have its velocity in a certain interval about the critical velocity, or else it either escapes the system or falls to the innermost critical orbit, or hits the emitting surface if that critical orbit lies within the radius  $R$ of that surface.

\section{Critical azimuthal velocity}

The first condition (\ref{nu0limit}) that a critical circular orbit exist with constant azimuthal velocity $\nu^{\hat\phi}_0$ can be restated in terms of the photon energy-momentum tensor. It simply requires that the azimuthal component of the photon azimuthal velocity as seen by the test particle vanish, which is equivalent to requiring that the azimuthal component of the energy-momentum tensor as seen by the test particle (namely the flux) vanish,
\beq
0=\sigma^{-1}\gamma^{-3} {\mathcal F}_{\rm (rad)}(U)^{\hat\phi}
=  T^{\hat 0\hat \phi} \nu^{\hat\phi}_0{}^2
  -(T^{\hat \phi\hat \phi}+T^{\hat 0\hat 0}) \nu^{\hat\phi}_0 
  +T^{\hat 0\hat \phi}
 \,.
\eeq 
When $T^{\hat0\hat\phi}=0$, this forces $\nu^{\hat\phi}_0=0$, but otherwise this condition is a quadratic equation for the critical velocity, namely
\beq
\label{fradphi_eq_02}
0=
T^{\hat 0\hat \phi}
( \nu^{\hat\phi}_0{}^2- W^{-1} \nu^{\hat\phi}_0 +1 )
=
T^{\hat 0\hat \phi}
(\nu^{\hat\phi}_0 -V) (\nu^{\hat\phi}_0 -V^{-1})
\,,
\eeq
where 
\beq
W= \frac{T^{\hat 0\hat \phi}}{T^{\hat 0\hat 0}+T^{\hat \phi\hat \phi}}\,,\
V= \frac12 W^{-1}  [1-(1-4W^2)^{1/2}]\,.
\eeq
The only physical root satisfying $|\nu^{\hat\phi}_0|<1$ is $\nu^{\hat\phi}_0=V$. Thus given a value of the critical radius for a given system, the critical azimuthal velocity (with the same sign as the azimuthal photon flux) follows immediately from this equation if one can evaluate the components of the radiation energy-momentum tensor at that radius.
Note that in the limit $|W|\to 0$ of small velocities, then $V\to W$, which is the slow rotation limit result noted by Miller and Lamb \cite{mil-lam3} in their Eq.~(18) taking into account finite size effects of the radiation source, and later investigated numerically by Oh, Kim and Lee and \cite{oh}.

If we re-examine the full equations for the radiation force, one sees the different contributions to the Poynting-Robertson effect for $T^{\hat0\hat\phi}\neq0$,
\begin{eqnarray}
\sigma^{-1}\gamma^{-3} {\mathcal F}_{\rm (rad)}(U)^{\hat r}
&=&
\gamma^{-2} (
T^{\hat 0\hat r} -T^{\hat r\hat\phi} \nu^{\hat\phi})
- \nu^{\hat r} P
\,,
\nonumber\\
\sigma^{-1}\gamma^{-3} {\mathcal F}_{\rm (rad)}(U)^{\hat\phi}
&=&
-T^{\hat0\hat\phi} (\nu^{\hat\phi}-V) (V^{-1}-\nu^{\hat\phi})
 - \nu^{\hat r} Q
\,,
\end{eqnarray}
where
\begin{eqnarray}
\fl\qquad
P
&=&
T^{\hat0\hat0}+ T^{\hat r\hat r} -2  T^{\hat0\hat\phi} \nu^{\hat\phi} +  (-T^{\hat r\hat r}+T^{\hat\phi\hat\phi})(\nu^{\hat\phi})^2
\,,
\nonumber\\
\fl\qquad
Q
&=&
[T^{\hat r\hat\phi}-2 T^{\hat0\hat r}  \nu^{\hat\phi}+ T^{\hat r\hat\phi}  (\nu^{\hat\phi})^2]
+[T^{\hat0\hat\phi}+ (T^{\hat r\hat r}-T^{\hat\phi\hat\phi})\nu^{\hat\phi}
   ] \nu^{\hat r}
-T^{\hat r\hat\phi}  (\nu^{\hat r})^2
\,.
\end{eqnarray}
Consider only outgoing photons.
The first term in the radial force is the flux seen by the test particle and responsible for the outward radiation pressure force,
which combines with the gravitational force contributions and the centrifugal force to rebalance the net radial force for circular orbits, while the second term is a radial drag force when $P>0$ as occurs in our case.  However, as noted earlier, this coefficient decreases roughly like $1/r^2$ at large distances and so may render the drag force ineffective in preventing some particles with sufficient initial energy from escaping to infinity.
The corresponding equation of motion then determines the critical radius of the radial equilibrium when that velocity goes to zero.
Since $T^{\hat0\hat\phi}  (V^{-1}-\nu^{\hat\phi})>0$,
the first term in the azimuthal force is an azimuthal drag towards the terminal azimuthal velocity $V$ (or towards $\nu^{\hat\phi}=0$ when $T^{\hat0\hat\phi}=0$), while the second term couples the radial velocity to this degree of freedom until the radial velocity goes to zero.

\section{Concluding remarks}

We have studied the behavior of test particles moving in a gravitational background of a black hole while subject to a Thomson-type interaction with a superimposed test radiation field (the Poynting-Robertson effect). In a previous article we considered a radiation flux outgoing in a purely radial direction with respect to the ZAMO family of observers in the equatorial plane of a Kerr background. We found that particles in motion in this plane which do not escape are attracted to a unique critical radius outside the horizon where they stay in radial equilibrium at rest with respect to the ZAMOs. In the present article, we have extended the problem to a coherent photon flux propagating in a general direction within the equatorial plane by allowing a non-zero photon angular momentum, which leads to an interesting interplay of gravitational dragging with the azimuthal drag exerted by the radiation. 
For an outgoing photon flux, bound particles end up in circular orbits with the same azimuthal velocity with respect to the ZAMOs as the photons.
However, the critical radius is not unique for sufficiently large values of the impact parameter $b$ as well as of the interaction parameter $A$.
An additional pair of critical orbits occur near the black hole, which remain relevant to a model of some massive object with a cutoff in the radius at its surface even if they occur inside that cutoff since the innermost critical orbit is an attractor causing particles to fall into the emitting surface unless they have special initial data to end up in the outermost critical orbit.
The gravitational dragging of course introduces an asymmetry into this system.

The present study is complementary to previous work on this problem taking into account the finite size of the radiation source, which leads to complicated integrals over the source in the rotating source case involving numerical ray tracing of the photons which arrive at the position of the test particle. This latter work is limited severely by the slow rotation condition on the radiation source. Our model allows some hint of strong rotation effects by ignoring the finite size of the radiation source.

The most natural directions for possible further refinement of this toy model include the choice of a more complex and astrophysically relevant radiation field,  a more sophisticated and realistic description of the particle-flux interaction, and possibly allowing for a more complicated structure for the test particle (higher multipole moments). Two improvements of the particle-flux interaction were studied by \cite{KeaneBS-01}, for example. They argued that at higher frequencies the interaction cross section is dominated by Compton scattering, which is not frequency independent as assumed here. The Compton scattering, in turn, tends to transfer energy to the test particle (the particle does not radiate all the acquired thermal energy away), thus effectively increasing its inertial mass. It would also be of interest to compare the results obtained for a test radiation field  with those obtained for the flux involved in the exact radiating solution of the Einstein equations due to Vaidya \cite{vaidya,vaidyakerr,lsm}.

\appendix

\section{Stability of the critical orbits}

For the general Kerr equatorial plane case considered here, the equilibrium solutions representing the critical circular orbits can be analyzed for their stability properties under small first order linear perturbations.
Let 
\beq
r=r_0\,, \quad 
\phi=\phi_0(\tau)\,, \quad
\nu=\nu_0\not=0\,, \quad
\alpha=\alpha_0\ (\cos\alpha_0=\pm1)
\eeq
be the parametric equation of an equilibrium solution, or symbolically $X^\alpha=X_0^\alpha$ ($\alpha=r,\phi,\nu,\alpha$). Recall that the sign $\cos\alpha_0=\pm1$ appearing here distinguishes the corotating and counter-rotating circular orbits. Note that this analysis will not check for stability against perturbations away from the equatorial plane.

Consider the linear perturbations of this solution $X^\alpha=X_0^\alpha+X_1^\alpha$, namely
\beq
\fl\quad
r=r_0+r_1(\tau)\ , \quad 
\phi=\phi_0(\tau)+\phi_1(\tau), \quad
\nu=\nu_0+\nu_1(\tau)\ , \quad
\alpha=\alpha_0+\alpha_1(\tau)\,,
\eeq
which leads to the following linear system of constant coefficient homogeneous linear differential equations
\beq
\frac{\rmd X_1^\alpha}{\rmd\tau}=C^\alpha{}_\beta X_1^\beta\,,
\eeq
which can easily be solved in terms of the eigenvalues and eigenvectors of the coefficient matrix. The real parts of all eigenvalues must be nonnegative for stability.
The explicit expressions for these coefficients in the Kerr case and their subsequent analysis are too complicated to reproduce here so we limit ourselves to the Schwarzschild case, where for $\nu_0\neq0$ the nonzero coefficients are
\begin{eqnarray}\fl
C^1{}_{4}&=&\pm\frac{\gamma_0\nu_0r_0\zeta_K}{\nu_K}\,,\quad
C^2{}_{1} = \mp\frac{\gamma_0\nu_0}{r_0^2}\,,\quad
C^2{}_{3} = \pm\frac{\gamma_0^3}{r_0}
\,,\nonumber\\\fl
C^3{}_{1}&=& \frac{\gamma_0^2\nu_0\zeta_K}{\gamma_K^2\nu_Kr_0}(\nu_0^2-\nu_K^2)[{\rm sgn}(\sin\beta_0)]
\,,\quad
C^3{}_{3} = \frac{\gamma_0^2\zeta_K}{\nu_K}(\nu_0^2-\nu_K^2)[{\rm sgn}(\sin\beta_0)]
\,, \nonumber\\\fl
C^3{}_{4}&=&\mp\frac{\gamma_0\nu_0^2\zeta_K}{\gamma_K^2\nu_K}\,,\quad
C^4{}_{1} = \mp\frac{\gamma_0^3\zeta_K}{\nu_K\nu_0r_0}[\nu_K^4+2\nu_K^2\nu_0^2(\nu_0^2-2)+\nu_0^2(2\nu_0^2-1)]
\,, \nonumber\\\fl
C^4{}_{3}&=&\pm\frac{2\gamma_0^3\zeta_K}{\gamma_K^2\nu_K}\,,\quad
C^4{}_{4} = \frac{2\gamma_0^2\zeta_K}{\nu_K}(\nu_0^2-\nu_K^2)[{\rm sgn}(\sin\beta_0)]\,.
\end{eqnarray} 
Here $\nu_K = \sqrt{{M}/{(r_0-2M)}}$ is the circular geodesic speed and $\zeta_K = \sqrt{M/r_0^3}$ is the corresponding coordinate time angular velocity.
The associated eigenvalues are 
\begin{eqnarray}\fl
\label{eigenvschw}
\lambda_0&=&0\,,\quad
\lambda_1 =\frac{\gamma_0^2\zeta_K}{\nu_K}(\nu_0^2-\nu_K^2)[{\rm sgn}(\sin\beta_0)]
\,, \nonumber\\\fl
\lambda_2&=&\lambda_1+i\Lambda\,,\quad
\lambda_3 =\lambda_1-i\Lambda\,,\quad 
\Lambda\equiv \gamma_0^2\frac{(\pm\nu_0)}{r_0}\sqrt{\nu_0^2-\nu_K^2+\frac{r_0^2\Omega_{\rm (ep)}^2}{\gamma_K^4\nu_K^2}}
\,,
\end{eqnarray}
where 
$\label{eq:Omegaep}
\Omega_{\rm (ep)}=\sqrt{M/r_0^3} \sqrt{(r_0-6M)/(r_0-3M)} 
$
is the corresponding proper time normalized version of the well known time coordinate epicyclic frequency governing the radial perturbations of circular geodesics.
Note that the last term in the square root expression appearing in $\Lambda$ can be rewritten as
\beq
\frac{r_0^2\Omega_{\rm (ep)}^2}{\gamma_K^4\nu_K^2}
=\left(\frac{\Omega_{\rm (ep)}}{\gamma_K\Omega_{\rm (orb)}}\right)^2=\frac{(r_0-3M)(r_0-6M)}{r_0(r_0-2M)}\ ,
\eeq
where $\Omega_{\rm (orb)}=\gamma_K\nu_K/{r_0} $ is the proper time orbital angular velocity of the geodesics.
Thus when $\nu_0=\nu_K$ one has $\lambda_1=0$, whereas $\lambda_2=-\lambda_3=\pm i\Omega_{\rm (ep)}$. 
Therefore, the eigenvalues all vanish for the circular geodesics at $r_0=6M$ and $\nu_0=1/2=\nu_K|_{r=6M}$. Finally observe that $\lambda_1\le0$ since
the outgoing/ingoing photon case (${\rm sgn}(\sin\beta_0)=\pm1$) correlates with $\nu_K$ being greater/less than $\nu_0$ when $A\neq0$.

While the above coefficient matrix is singular for  $\nu_0=0$ (a case for which therefore ${\rm sgn}(\sin\beta_0)=1$), it turns out that the limiting values of the eigenvalue formulas from above remain valid,
with $\lambda_0=0$ and $\lambda_1=\lambda_2=\lambda_3=-\nu_K\zeta_K$ all negative, so this case is always stable. 
However, to show this in detail it is convenient to perform the previous linearization around the equilibrium solution using the signed radial and azimuthal linear velocities,
 i.e., $\nu^{\hat r}=\nu\sin\alpha$ and $\nu^{\hat \phi}=\nu\cos\alpha$ respectively. Then
\beq
\fl\qquad
r=r_0+r_1(\tau)\,, \quad 
\phi=\phi_0+\phi_1(\tau)\,, \quad
\nu^{\hat r}=\nu^{\hat r}_1(\tau)\,, \quad
\nu^{\hat \phi}=\nu^{\hat \phi}_1(\tau)
\eeq
leads to the linearized differential equations
\begin{eqnarray}
&\frac{\rmd r_1}{\rmd\tau}=\frac{r_0\zeta_K}{\nu_K}\nu^{\hat r}_1\,, \qquad
&\frac{\rmd \nu^{\hat r}_1}{\rmd\tau}=-\nu_K\zeta_K\left(\nu_K^2\frac{r_1}{r_0}+2\nu^{\hat r}_1\right)\,, \nonumber\\
&\frac{\rmd \phi_1}{\rmd\tau}=\frac{\nu^{\hat \phi}_1}{r_0}\,, \qquad
&\frac{\rmd \nu^{\hat \phi}_1}{\rmd\tau}=-\nu_K\zeta_K\nu^{\hat \phi}_1
\end{eqnarray}
whose coefficient matrix has exactly the eigenvalues stated above.

For the two geodesic curves $A=0$ (thick black curves) representing $b$ versus $r_0$ for the circular geodesic orbits shown in the lower graph of Fig.~\ref{fig:radiusnu}, the single local extremum at $r_0=6M$ divides the stable orbits at larger $r_0$ from the unstable orbits at smaller $r_0$.  This is the so called ``last stable circular orbit" as one approaches the black hole horizon.
For nonzero values of $A$, one finds similarly that the unstable critical orbits lie entirely within the simple closed curves of local extrema of the family of $b$ versus $r_0$ curves of constant $A$ shown as the shaded region in Fig.~\ref{fig:radiusnu}: the two bounding curves for this region are $\lambda_2 =0$ and $\lambda_3 =0$. For a fixed value of $A$ whose curve passes through this region and for each value of $|b|$ between its minimum and maximum value on the boundary of this shaded region, there are 3 critical orbits, and the intermediate one whose radius lies within this region is unstable, while the other two which lie outside are stable, as is the single critical orbit which occurs outside this region for values of $b$ outside this range.

To establish this, consider the equilibrium condition (\ref{AMN}), i.e.,
\beq\label{AMN2}\fl\qquad
\frac{A}{M N}
={\rm sgn}(\sin\beta_0) \displaystyle \frac{1-\displaystyle\frac{ b^2}{M r_0 }\left(1-\displaystyle\frac{2M}{r_0}\right)^2 }
{\left[ 1-\displaystyle\frac{b^2}{r_0^2}\left(1-\displaystyle\frac{2M}{r_0}\right) \right]^{3/2}}
   \,,
\eeq
which in principle can be solved for $b=\tilde b(r_0)$.
One can easily show by implicit differentiation when $\nu_0\neq0$ that
\begin{eqnarray}
\frac{\rmd\tilde b}{\rmd r_0}&=&\lambda_2\lambda_3\frac{r_0^8}{\tilde b\gamma_0^4(r_0-2M)}\frac{1}{[(r_0-2M)^2\tilde b^2+r_0^3(2r_0-7M)]}\nonumber\\
&=&\lambda_2\lambda_3\frac{\pm\nu_K^3}{r_0\nu_0\gamma_0^2\zeta_K^3}\frac{1}{(\nu_0^2-3\nu_K^2+2)}\ ,
\end{eqnarray}
taking into account that 
$\nu_0=\pm (b/r_0)(1-2M/r_0)^{1/2}$.
Therefore the boundary of the instability region, which is enclosed by the two curves $\lambda_2=0$ and $\lambda_3=0$, coincides exactly with the curve of critical points of the constant $A$ curves $b$ versus $r_0$.

Turning on the Kerr rotation parameter $a$, 
the generalization of the present analysis is straightforward though it involves much more complicated formulas.
Again the region of unstable critical orbits is a simple closed curve joining together the critical points of the constant $A$ curves in the $b$ versus $r_0$ diagram. This is the pair of shaded regions shown in Figs.~\ref{fig:radiusnukerr1} and \ref{fig:radiusbkerr1}, which are asymmetrically deformed by increasing values of the rotation parameter $a$ until the upper region disappears at the outer horizon where the corotating circular geodesic radius squeezes to that outer horizon, leaving only one stable critical orbit radius for all values of $b\ge0$ of the corotating outgoing photon case.

\section*{Acknowledgements}

DB and RTJ thank ICRANet for its continued support.
OS thanks the Istituto per le Applicazioni del Calcolo ``M. Picone'' for its hospitality and Czech projects
GACR-202/09/0772, MSM0021610860 and LC06014 for support.

\end{document}